\pdfoutput=1 
\newcommand*{\ATLASLATEXPATH}{}
\documentclass[cernpreprint, USenglish, texlive=2016]{\ATLASLATEXPATH atlasdoc}
\usepackage{\ATLASLATEXPATH atlaspackage}
\usepackage{textgreek}
\usepackage{multirow}
\usepackage{wasysym}
\usepackage{\ATLASLATEXPATH atlasbiblatex}
\usepackage{\ATLASLATEXPATH atlasphysics}
\usepackage{EXOT-2017-08-defs}
\AtlasTitle{
	Search for pairs of highly collimated photon-jets in $pp$ collisions at $\sqrt{s}$ = 13~\TeV\ with the ATLAS detector	
}

\AtlasRefCode{EXOT-2017-08}

\AtlasJournal{PRD}

\AtlasJournalRef{Phys. Rev. D 99 (2019) 012008}

\AtlasDOI{10.1103/PhysRevD.99.012008}

\AtlasAbstract{%
Results of a search for the pair production of photon-jets---collimated groupings of photons---in the ATLAS detector at the Large Hadron Collider are reported.
Highly collimated photon-jets can arise from the decay of new, highly boosted particles that can decay to multiple photons collimated enough to be identified in the electromagnetic calorimeter as a single, photonlike energy cluster.
Data from proton-proton collisions at a center-of-mass energy of 13~\TeV, corresponding to an integrated luminosity of 36.7~fb$^{-1}$, were collected in 2015 and 2016.
Candidate photon-jet pair production events are selected from those containing two reconstructed photons using a set of identification criteria much less stringent than that typically used for the selection of photons, with additional criteria applied to provide improved sensitivity to photon-jets.
Narrow excesses in the reconstructed diphoton mass spectra are searched for.
The observed mass spectra are consistent with the Standard Model background expectation.
The results are interpreted in the context of a model containing a new, high-mass scalar particle with narrow width, $X$, that decays into pairs of photon-jets via new, light particles, $a$.
Upper limits are placed on the cross section times the product of branching ratios $\sigma \times \mathcal{B}(X \rightarrow aa) \times \mathcal{B}(a \rightarrow \gamma \gamma)^{2}$ for $200~\GeV< m_{X} <2~\TeV$ and for ranges of $ m_a $ from a lower mass of $ 100~\MeV $ up to between 2 and $ 10~\GeV $, depending upon $ m_X $.
Upper limits are also placed on $\sigma \times \mathcal{B}(X \rightarrow aa) \times \mathcal {B}(a \rightarrow 3\pi^{0})^{2}$ for the same range of $ m_X $ and for ranges of $ m_a $ from a lower mass of $ 500~\MeV $ up to between 2 and $ 10~\GeV $.
}

\AtlasCoverSupportingNote{Support note}{https://cds.cern.ch/record/2259709}
\AtlasCoverCommentsDeadline{25 July 2018}

\AtlasCoverAnalysisTeam{James Beacham, Yuya Kano, Yasuyuki Okumura, Elisabeth Petit}

\AtlasCoverEdBoardMember{Nicolas Berger~(chair), Daniel Hayden, Ruggero Turra}
\AtlasCoverEgroupEditors{atlas-exot-2017-08-editors@cern.ch}

\AtlasCoverEgroupEdBoard{atlas-exot-2017-08-editorial-board@cern.ch}

\PreprintIdNumber{CERN-EP-2018-143}
\hypersetup{pdftitle={ATLAS document},pdfauthor={The ATLAS Collaboration}}

\begin{document}

\maketitle

\tableofcontents

\newpage

\section{Introduction} \label{sec:Intro}
The quest for new particles at the Large Hadron Collider (LHC) at CERN has been greatly rewarded by closely examining collision events that contain photons in the final state.
Despite the relatively small branching ratio predicted for the process in the Standard Model (SM), the decay of the Higgs boson into two photons is readily identifiable due to the good energy resolution of the electromagnetic (EM) calorimeters of the ATLAS~\cite{atlas-detector} and CMS~\cite{Chatrchyan:2008aa} detectors and the relatively small backgrounds in final states with only photons. The search for this process was one of the main methods by which the Higgs boson was observed~\cite{Aad:2012tfa,Chatrchyan:2012ufa}.
Moreover, the establishment of a wide range of results that so far are consistent with the SM at the LHC at a center-of-mass energy of 13~\TeV\ motivates a renewed focus on searches for new physics that target general experimental signatures, including nonstandard photon signatures, rather than specific signal models. 
In many beyond the Standard Model (BSM) theories~\cite{Dobrescu:2000jt,Toro:2012sv,Knapen:2015dap,Agrawal:2015dbf,Chang:2015sdy,Aparicio:2016iwr,Dasgupta:2016wxw,Domingo:2016unq,Chiang:2016eav}, new scalar, pseudoscalar or vector gauge bosons can decay into photon-only final states that lead to collimated groupings of photons (``photon-jets''~\cite{Ellis:2012zp,Ellis:2012sd}). 
In some cases, the Lorentz boost of the new particles is large enough to lead to an opening angle between the trajectories of the final-state photons that is smaller than or comparable to the angular size of an energy cluster in the EM calorimeter corresponding to a single photon, resulting in highly collimated photon-jets.
Such boosted particles arise, for example, when a high-mass particle produced in the proton-proton collision decays into intermediate particles, with much lower masses, that subsequently decay into photons.
Thus, events selected to contain two, well-separated, reconstructed photons can be used to search for pairs of highly collimated photon-jets resulting from BSM particle decays.

A search for highly collimated photon-jets using 36.7 fb$^{-1}$ of LHC proton-proton collision data collected by the ATLAS detector in 2015 and 2016 at a center-of-mass energy of 13~\TeV\ is presented.
Candidate photon-jet pair production events are selected from those containing two reconstructed photons (denoted ``$ \gamma_R $''), using a set of identification criteria much less stringent than that typically used for the selection of photons, and with additional criteria applied to provide improved sensitivity to photon-jets.
Narrow excesses are searched for in the spectra of the reconstructed diphoton mass $ \mgg $.

\begin{figure}[h!]
  \centering
  \subfloat{
  \includegraphics[width=0.45\columnwidth]{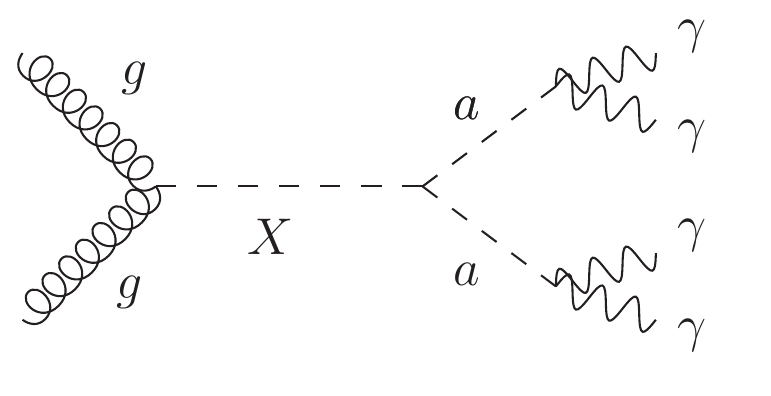}
  \label{fig:hToAATo4Ph}
  }
  \hskip 10mm
  \subfloat{
  \includegraphics[width=0.45\columnwidth]{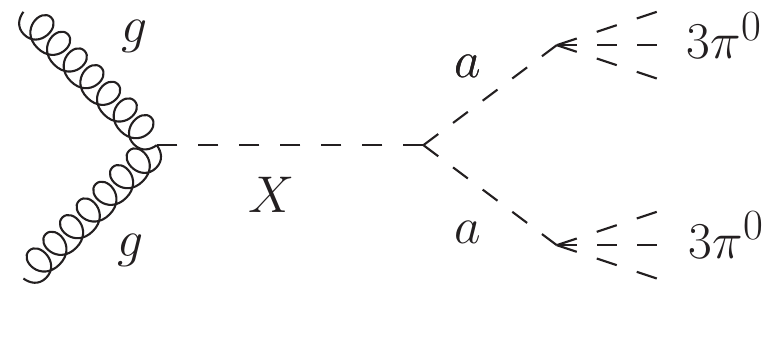}
  \label{fig:hToAATo6pion}
  }
  \caption{Diagrams for BSM scenarios that result in events with pairs of photon-jets in the final state.
}
  \label{fig:BSMdiagrams}
\end{figure}

The results of the search are interpreted in the context of a benchmark BSM scenario involving a high-mass, narrow-width scalar particle, $X$, with mass $ m_X > 200~\GeV $, originating from the gluon-gluon fusion process and that can decay into a pair of intermediate particles with spin 0, $a$, as shown in Figure~\ref{fig:BSMdiagrams}.
The $a$ particle can in general decay to several final states, but here is restricted to decay either into a pair of photons, via $X \rightarrow aa \rightarrow 4\gamma$, or into three neutral pions, via $X \rightarrow aa \rightarrow 6\pi^{0} \rightarrow 12\gamma$, yielding events containing a pair of photon-jets of either low or high multiplicity; the result is interpreted for both cases.
Because the search is performed using events that contain two calorimeter deposits that are initially loosely identified as individual photons, the search is sensitive to the parameter region in which the $ a $ particle is highly boosted, $  m_a < 0.01 \times  m_X $.

\section{ATLAS detector} \label{sec:Detector}
\interfootnotelinepenalty=10000
\newcommand{\AtlasCoordFootnote}{%
ATLAS uses a right-handed coordinate system with its origin at the nominal interaction point in the center of the detector and the $z$-axis along the beam pipe.
The $x$-axis points from the interaction point to the center of the LHC ring,
and the $y$-axis points upwards.
Cylindrical coordinates $(r,\phi)$ are used in the transverse plane, 
$\phi$ being the azimuthal angle around the $z$-axis.
The pseudorapidity is defined in terms of the polar angle $\theta$ as $\eta = -\ln \tan(\theta/2)$.
Angular distance is measured in units of $\Delta R \equiv \sqrt{(\Delta\eta)^{2} + (\Delta\phi)^{2}}$.}

The ATLAS detector~\cite{atlas-detector} is a multipurpose detector with a forward-backward symmetric cylindrical geometry.\footnote{\AtlasCoordFootnote}
The detector covers nearly the entire solid angle around the collision point.
It consists of an inner tracking detector surrounded by a thin superconducting solenoid, EM and hadronic calorimeters, and a muon spectrometer incorporating three large superconducting toroid magnets.  

The inner-detector system is immersed in a \SI{2}{\tesla} axial magnetic field and provides charged-particle tracking in the range $|\eta| < 2.5$.
The high-granularity silicon pixel detector covers the vertex region.
The innermost layer of the pixel detector, the insertable B-layer~\cite{ATLAS-TDR-19}, was installed between Run 1 and Run 2 of the LHC\@.
The pixel detector typically provides four measurements per track.
It is followed by the silicon microstrip tracker that normally provides four two-dimensional measurement points per track.  These silicon detectors are complemented by the transition radiation tracker, which enables radially extended track reconstruction up to $|\eta| = 2.0$.  The transition radiation tracker also provides electron identification information based on the fraction of hits (typically 30 in total) above a higher energy-deposit threshold corresponding to transition radiation.

The calorimeter system covers the pseudorapidity range $|\eta| < 4.9$.  
Within the region $|\eta|< 3.2$, EM calorimetry is provided by a high-granularity lead/liquid-argon (LAr) EM calorimeter.
The EM calorimeter is divided into a barrel section covering $|\eta|<1.475$ and two endcap sections covering $1.375<|\eta|<3.2$. 
For $|\eta|<2.5$, the EM calorimeter is composed of three sampling layers in the longitudinal direction of shower depth. 
The first layer is segmented into high-granularity strips in the $\eta$ direction, with a typical cell size of $\Delta\eta\times\Delta\phi=0.003\times0.1$ for the ranges $|\eta|<1.4$ and $1.5<|\eta|<2.4$, and a coarser cell size of $\Delta\eta\times\Delta\phi=0.025\times0.1$ for other regions. This fine granularity in the $\eta$ direction allows identification of events with two overlapping showers originating from the decays of neutral hadrons in hadronic jets, mostly $\pi^0\to\gamma\gamma$ decays. The second layer has a cell size of $\Delta\eta\times\Delta\phi=0.025\times0.025$. This layer collects most of the energy deposited in the calorimeter by photon and electron showers. The third layer is used to correct for energy leakage beyond the EM calorimeter from high-energy showers.
The thicknesses of the first, second, and third layers at $ \eta=0 $ are 4.3 radiation lengths ($X_0$),  $16~X_0$, and $2~X_0$, respectively, and they vary with the pseudorapidity range~\cite{atlas-detector}.
Placed in front of these layers, an additional thin LAr presampler layer covering $|\eta| < 1.8$ is used to correct for energy loss in material upstream of the calorimeters.  Hadronic calorimetry is provided by the steel/scintillator-tile calorimeter, segmented into three barrel structures within $|\eta| < 1.7$, and two copper/LAr hadronic endcap calorimeters.  The solid angle coverage is completed with forward copper/LAr and tungsten/LAr calorimeter modules optimized for EM and hadronic measurements respectively.

A two-level trigger system, the first level implemented in custom hardware followed by a software-based level, is used to reduce the event rate to about 1 kHz for offline storage.

\section{Photon-jet signal characteristics} \label{sec:SigChar}
Photon-jets, defined as collimated groupings of photons, can arise from decays of particles that are highly boosted as a result of themselves being the decay products of higher-mass particles.
For the benchmark BSM scenario considered here, the extent to which photons from decays of $ a $ particles are collimated depends on the ratio of the masses of the $X$ and $a$, particularly in the case where the $ X $ particle is produced with a momentum significantly less than its mass.

For large values of the ratio $m_{a}$/$m_{X}$, the boost of the $a$ is small enough to yield more than two individual photons, well separated and isolated, that can be identified in the detector.
In this regime, a general search for new phenomena in events with at least three isolated photons, using a three-photon trigger, was performed by ATLAS at 8~\TeV~\cite{Aad:2015bua}.
This search was sensitive to cases where the angular separation between photons was large, for $\dR \gtrsim$ 0.3, which corresponds to $ m_a/m_X \gtrsim 0.08 $ for the benchmark signal scenario.
For slightly smaller values of the ratio $m_{a}$/$m_{X}$, the individual final-state photons appear too close together in the detector and fail isolation criteria, limiting the sensitivity of the 8~\TeV\ ATLAS search in this regime.

For very small values of the ratio $ m_a/m_X $, the boost of the $ a $ is large enough to lead to angular separations between the final-state photons of $\dR \lesssim 0.04$, which is approximately the same size as a standard single-photon energy cluster in the ATLAS EM calorimeter.
In this case, existing triggers cannot distinguish a calorimeter energy deposit resulting from highly collimated photons from that of a single photon.  
Thus, diphotonlike events can be used as a starting point for a search for highly collimated photon-jets, and the sensitivity to this region of the photon-jet parameter space can be increased by placing criteria on the shape of the shower in the EM calorimeter in addition to those applied in the trigger.
This analysis presents a search for highly collimated photon-jets that is sensitive to a wide mass range for the parent $X$ particle, $m_{X} >$ 200~\GeV, and for $m_{a}/m_X < 0.01$ in the benchmark signal scenario.

\begin{figure}[h!]
	\centering
	\subfloat{\includegraphics[width=0.6\textwidth]{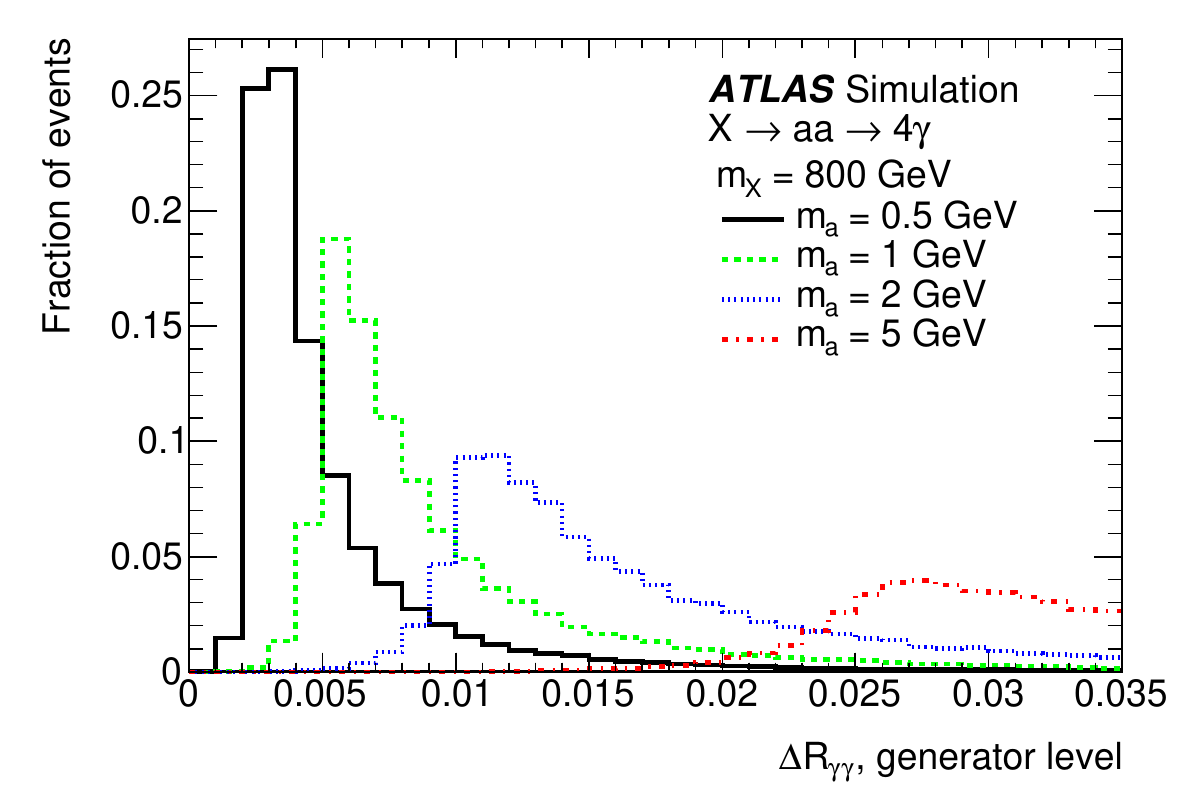}}
	\caption{The distribution of $ \dR $, the angular separation between two photons that are reconstructed as a single  photon-jet in the ATLAS detector, for the benchmark signal scenario for the process $ X \to aa \to 4\gamma $, using simulated signal samples at generator level. The distribution has a peak at $\sim 4\times m_a/m_X $ and a long tail on the right side. 
	For the values of $ m_X $ and $ m_a $ presented in the figure, $ \dR $ is smaller than or comparable in size to an EM cluster.}
	\label{fig:truthDeltaR}
\end{figure}

For this benchmark scenario, for the process $X \to aa \to 4\gamma$, the distribution of $\dR$ is shown in Figure~\ref{fig:truthDeltaR}.
Due to the kinematics of boosted particles, $\dR$ has a maximum at a value of $ 2/\gamma $, where $ \gamma $ is the Lorentz factor of the $ a $ particle, $ \gamma=E_a/m_a $.
When the $ X $ particle is produced nearly at rest, since the energy of the $ a $ particle has a median value of $ E_a \sim m_X/2 $, the distribution of $ \dR $ has a maximum at $\sim 4\times m_a / m_X $.
The approximate proportionality of the angular spread of photons within the photon-jet to $m_a / m_X$ holds for photon-jets in general, including those with larger photon multiplicity resulting from processes such as $ X\to aa\to 6\pi^0 $.
Since the two different final states of the benchmark scenario are similar, some parts of the descriptions in the following sections are only mentioned for the $ X\to aa\to 4\gamma $ decay to avoid repetition, although they apply to the $ X\to aa\to 6\pi^0 $ decay as well.

For values of the ratio $m_{a}$/$m_{X}$ greater than 0.01, the final-state photons are separated enough to lead to a relatively large cluster of energy in the calorimeter, and such events do not satisfy the isolation criteria or the initial loose identification of photons at the trigger level.
The signal selection efficiency for the present analysis in this $m_a / m_X > 0.01$ region is lower than 4\%, and so no attempt is made to search in this regime.
There is therefore an intermediate region, $ 0.01 <m_a/m_X \lesssim 0.08 $, which is covered by neither this search nor the previous search for three-photon final states at 8~\TeV.

\section{Event samples} \label{sec:EventSamples}
The data sample used for this search corresponds to an integrated luminosity of 36.7~fb$^{-1}$ (after applying data-quality requirements), collected under normal data-taking conditions for $pp$ collisions during 2015 and 2016 at a center-of-mass energy of $\sqrt{s} =$ 13~\TeV.
The data were selected using an unprescaled trigger that filters events with two energy deposits in the EM calorimeter that satisfy trigger-level loose photon identification criteria with transverse energy values of $E_\text{T,1} >$ 35~\GeV\ and $E_\text{T,2} >$ 25~\GeV.

Samples of the benchmark signal scenario with two different final states, $ X\to aa\to 4\gamma$ and $ X\to aa\to 6\pi^0 $, were simulated.
For the production of the $X$ via gluon-gluon fusion, \MGMCatNLO~\cite{Alwall:2014hca} Version 2.3.3, at next-to-leading order (NLO) in quantum chromodynamics (QCD) with the NNPDF30NLO parton distribution function (PDF) set~\cite{Ball:2014uwa}, was used. 
For the subsequent decay of the $X$ into $aa$ and into the photon-jet final states, \PYTHIA{8}~\cite{Sjostrand:2007gs} Version 8.210, with the A14 set of tuned parameters~\cite{ATL-PHYS-PUB-2014-021}, was used, as well as for the parton-shower and hadronization simulation of initial state radiation jets.
The samples were produced using a narrow-width approximation (NWA) approach with the resonance widths of the $X$ and $a$ set to 4~\MeV\ and 1~\MeV, respectively.
Samples were simulated for mass ranges of $ 200~\GeV < m_{X} < 2000~\GeV $ and $ 0.1~\GeV < m_{a} < 0.01 \times m_{X}$.

The nonresonant production of diphoton events in the SM is the dominant background source for this analysis, and these events were simulated with \SHERPAV{2.1.1}~\cite{Gleisberg:2008ta}.
Matrix elements were calculated with up to two additional partons at leading order (LO) in QCD and merged with the \SHERPA\ parton-shower simulation~\cite{Schumann:2007mg} using the ME+PS@LO prescription~\cite{Hoeche:2009rj}.
The CT10 PDF set~\cite{Lai:2010vv} was used in conjunction with a dedicated parton-shower tune of \SHERPA.
These samples are used to validate the background modeling based on analytic functions (described in Section~\ref{sec:Bkg_Modelling}).
Simulated samples of the reducible SM background consisting of one photon and one hadronic jet from the hard process were also generated with \SHERPAV{2.1.1}---using the same PDF set, parton-shower tune, and merging prescription as for the diphoton sample---with matrix elements calculated at LO with up to four additional partons.
These samples are used for optimizing the search strategy described in Section~\ref{sec:ObjAndEvtSel}.

Additional interactions in the same or nearby bunch crossings (pileup) were simulated using Pythia 8.186~\cite{Sjostrand:2007gs} using the A2 set of tuned parameters~\cite{ATL-PHYS-PUB-2012-003} and the MSTW2008LO PDF~\cite{Martin:2009iq} set and overlayed on the simulated signal and SM background events.
All simulated event samples were produced using the ATLAS simulation infrastructure~\cite{SOFT-2010-01}, using the full \GEANT4~\cite{Agostinelli:2002hh} simulation of the ATLAS detector.  Simulated events were then reconstructed with the same software as used for the data.

\section{Object and event selection} \label{sec:ObjAndEvtSel}
This analysis selects events containing at least two reconstructed photons, obtained from a diphoton trigger, and then searches for pair-produced photon-jets.
This is accomplished by applying additional selection criteria and scanning for deviations from the expected background in the $\mgg$ spectrum, defined as the distribution of the mass values of the two reconstructed photons, which would correspond to the mass of the high-mass particle $m_X$ in the case of a signal event.
No attempt is made to reconstruct the mass of the $a$ in the process $X \rightarrow aa \rightarrow$~photon-jets (although specifics of the $a$ are taken into account in several parameters of the signal modeling, which is detailed in Section~\ref{sec:SignalModelling}).

\subsection{Initial event selection with two loose photons} \label{sec:ObjAndEvtSel_BaselineEventSelection}

Reconstructed photons are obtained from clusters of energy deposited in the EM calorimeter~\cite{Aaboud:2016yuq}.
In the barrel section a cluster size of $ 3\times7 $ cells in the middle layer is used (equivalent to an area of size $ \Delta \eta \times \Delta \phi = 0.075 \times 0.175 $), while a cluster of $ 5\times 5 $ cells in the middle layer is used in the endcap sections (equivalent to an area of $ \Delta \eta \times \Delta \phi = 0.125 \times 0.125 $).
Reconstructed photons are required to match photon objects calculated at the trigger level, within the separation of $ \Delta R < 0.07 $, and may have associated tracks and conversion vertices reconstructed in the inner detector.

The two leading reconstructed photons are required to be within the fiducial calorimeter region of $|\eta| < 2.37$, excluding the transition region at $1.37 < |\eta| < 1.52$ between the barrel and endcap calorimeters.
The criterion $ E_\text{T,1} > 0.4 \times \mgg $ is applied to the leading reconstructed photon, and $ E_\text{T,2} > 0.3 \times \mgg $ to the subleading reconstructed photon.
These criteria increase the sensitivity to photon-jet pairs from a scalar resonance, since such candidate signal events tend to contain photons with larger $ \ET/\mgg $ ratios compared with those from background events dominated by $t$-channel processes~\cite{Aaboud:2016tru}.
Only events with $ \mgg > 175~\GeV $ are selected for further analysis.

The two leading reconstructed photons are required to be isolated from other calorimeter energy deposits and from nearby tracks not associated with the photon.
The calorimeter isolation variable $ \ET^{\text{iso}} $ is defined as the sum of energy deposits in the calorimeter in a cone of size $ \Delta R=0.4 $ around the barycenter of the photon cluster (excluding the energy associated with the photon cluster) minus $ 0.022\times \ET $.
This cone energy is corrected for the leakage of the photon energy from the photon cluster and for the effects of pileup~\cite{PERF-2013-05}.
The calorimeter isolation variable is required to satisfy $ \ET^{\text{iso}} < 2.45~\GeV $. 
The track isolation variable $ \pT^\text{iso} $ is defined as the scalar sum of the transverse momenta of tracks not associated with the photon in a cone of size $ \Delta R=0.2 $ around the barycenter of the photon cluster.
It is required to satisfy $ \pT^{\text{iso}} < 0.05\times \ET $.

\subsection{Optimized photon selection for photon-jet signatures} \label{sec:ObjAndEvtSel_CustomizedPhotonIdentification}

Photon identification in ATLAS~\cite{Aaboud:2016yuq} is based on a set of requirements placed on several discriminating variables that characterize the shower development in the calorimeter (``shower shapes''), defined to reject the background from hadronic jets misidentified as photons. 
Nine discriminating variables are defined, and they are described in detail in Table~1 of Ref.~\cite{Aaboud:2016yuq}.
One variable quantifies the shower leakage fraction in the hadronic calorimeter, and three variables quantify the lateral shower development in the EM calorimeter second layer. 
The other five variables quantify the lateral shower development in the finely segmented strips of the first layer, and two of them are utilized to identify photon candidates with two separate local energy maxima in the fine strips, which are characteristic of neutral hadron decays in hadronic jets, primarily from $\pi^0\to\gamma\gamma$.

Several reference selections are defined, including those referred to as ``loose'' and ``tight''. The loose selection is based only on shower shapes in the second layer of the EM calorimeter and on the leakage in the hadronic calorimeter, and is used by the photon triggers, including the diphoton trigger used for the collection of the data sample for this search.
The tight selection is based on all nine variables and is used for the standard photon identification in ATLAS, but is not used in this search. 
The criteria for the tight selection change as a function of the $\eta$ values of the reconstructed photons, to account for the calorimeter geometry and effects from the material upstream of the calorimeter, and are separately optimized for reconstructed photons with and without an associated conversion vertex to increase the photon identification efficiency.

In this search, both reconstructed photons are required to fulfill the ``loose$^\prime$'' selection. 
This selection is defined by removing requirements on all five variables quantifying the shower development in the finely segmented strip layer of the calorimeter ($ w_{s\ 3}, w_{s\ \text{tot}}, F_{\text{side}}, \Delta E$, and $ E_{\text{ratio}} $, defined in Table 1 of Ref.~\cite{Aaboud:2016yuq}), with respect to the standard tight selection.
The requirements on the other four variables ($ R_\text{had},  R_\eta, w_{\eta 2} $, and $ R_\phi $) remain the same as for the standard tight selection.
By definition, the loose$^\prime$ is an intermediate selection between loose and tight.
Based on simulated samples of signal and SM background processes, this loose$^\prime$ selection provides good sensitivity to photon-jet signals.
This is explained by the fact that energy clusters of photon-jets exhibit multiple local energy maxima in the fine strip layer, since the angular separation of photons constituting the photon-jet can be larger than the segmentation of the strips, depending on the mass parameters $m_X$ and $m_a$ of the benchmark signal scenario, as seen in Figure~\ref{fig:truthDeltaR}.
For signal mass values $ 0.003<m_a/m_X<0.006 $ and $ m_X>200~\GeV $, the total selection efficiency is less than 5\% when the standard tight selection is applied, in addition to the selection criteria described in Section~\ref{sec:ObjAndEvtSel_BaselineEventSelection}, and this increases to 20\%--50\% with the loose$^\prime$ selection. 
Comparing the two selection criteria, an increase in the overall event yield of roughly 30\% is observed with the loose$^\prime$ selection.
Thus, the analysis sensitivity to photon-jet signals is increased by the use of the loose$^\prime$ selection, rather than the standard tight selection.

Additionally, the choice of loose$^\prime$ allows the definition of a set of ``not loose$^\prime$'' criteria (i.e., where at least one of the two reconstructed photons fails the loose$^\prime$ selection) that is used to define the control regions for the evaluation of the background composition, as described in Section~\ref{sec:Bkg_Modelling}.

\subsection{Categorization of events by the shower shape variable $\Delta E$} \label{sec:ObjAndEvtSel_Categorization}

After the preselection of events with two leading reconstructed photons satisfying the isolation and loose$^\prime$ identification criteria described in the previous sections, the final signal region is defined by dividing the events into two orthogonal categories based on the value of the calorimeter variable $\Delta E$ for the reconstructed photons.
The quantity $ \Delta E $ corresponds to a shower shape variable based on information in the first layer of the EM calorimeter, and quantifies the relative size of multiple, individual energy deposits that may be contained within a single energy cluster.

It is defined as 
\begin{equation}
	\Delta E = E_\text{2nd max}^{S1} - E_\text{min}^{S1} \nonumber
\end{equation}
where $ E_\text{2nd max}^{S1} $ is the energy of the strip cell with the second-largest energy, and $ E_\text{min}^{S1} $ is the energy in the strip cell with the lowest energy located between the strips with the largest and the second-largest energy.
If the strip cells with the largest and the second-largest energy are located next to each other, or if there is no second-largest energy strip, then $ \Delta E = 0 $.
This variable is useful for identifying the $ \pi^0\to\gamma\gamma $ process, prevalent in hadronic jets, which leaves a characteristic signature in the first layer of the EM calorimeter that often yields two peaks in the $ \eta $ direction, resulting in large $ \Delta E $ values.
When the photon-jet signals from decays such as $ a\to\gamma\gamma $ and $ a\to3\pi^0\to6\gamma $ have angular separation of photons larger than the segmentation of the first layer of the EM calorimeter, they leave signatures in the calorimeter similar to $ \pi^0\to\gamma\gamma $ events. Thus, the variable $ \Delta E $ is used to effectively select photon-jet signals.

The categorization by $ \Delta E $ is as follows:
\begin{itemize}
	\item \textbf{Low-$\Delta E$ category:} both reconstructed photons are required to have values of $ \Delta E $ below given thresholds. This requirement corresponds to reconstructed photons with a signature in the fine strip layer similar to that of single photons.
	\item \textbf{High-$\Delta E$ category:} at least one of the two leading reconstructed photons is required to have a value of $ \Delta E $ above a given threshold. This requirement corresponds to events containing reconstructed photons which have a $\pi^0$-like signature.
\end{itemize}

The thresholds for the value of $ \Delta E $ used to determine whether an event appears in either the low- or high-$\Delta E$ category are the same as those used in the standard tight photon selection.
These thresholds range from 100 to $ 500~\MeV $, depending on the photon $ \eta $ and whether there are associated tracks or conversion vertices.

The high-$ \Delta E $ category is found to have a significantly better signal-to-background ratio compared with the low-$ \Delta E $ category, since reconstructed photons with large $ \Delta E $ values typically correspond to photon-jets with a larger angular spread among the constituent photons. 
The high-$ \Delta E $ criterion also effectively reduces the contribution of single photons, which tend to have small $ \Delta E $ values.
Hadronic jets from SM processes containing $ \pi^0\to\gamma\gamma $ decays are likely to fall into the high-$ \Delta  E$ category, but the contribution of these events is small due to the isolation requirements.
This leads to lower expected background event yields in the high-$ \Delta E $ category, resulting in better signal-to-background ratios compared with the low-$ \Delta E $ category.
The number of events observed in each category for different ranges of $\mgg$ is shown in Table~\ref{tab:dataYields}. 
Although overall the ratio of signal-to-background is lower for the low-$\Delta E$ category, it still provides increased sensitivity to photon-jet signals with smaller angular separation, and so both categories are used in this search.

\sisetup{scientific-notation=true, round-mode=figures, round-precision=2}
\begin{table}[h!]
	\begin{center}
		\caption{ Number of events observed in the two categories for different $\mgg$ ranges. }
		\begin{tabular}{l|cccc}
			\hline
			     $ \mgg $ range       &  175--400~\GeV  & 400--600~\GeV  & 600--800~\GeV &  $ > $800~\GeV   \\ \hline
			Low-$\Delta E$ category  & $5.3\times10^4  $ & $ 2.5\times10^3 $ & $ 5.2\times10^2 $ & $ 2.3\times10^2 $ \\
			High-$\Delta E$ category & $ 9.8\times10^3 $ & $ 3.5\times10^2 $ & $ 5.2\times10^1 $  & $ 2.1\times10^1 $  \\ \hline
		\end{tabular}
		\label{tab:dataYields}
	\end{center}
\end{table}

\subsection{Summary of the selection} \label{sec:ObjAndEvtSel_ChoiceDeltaE}

\begin{figure}[h!] 
	\centering
	\subfloat[]{\includegraphics[width=0.48\textwidth]{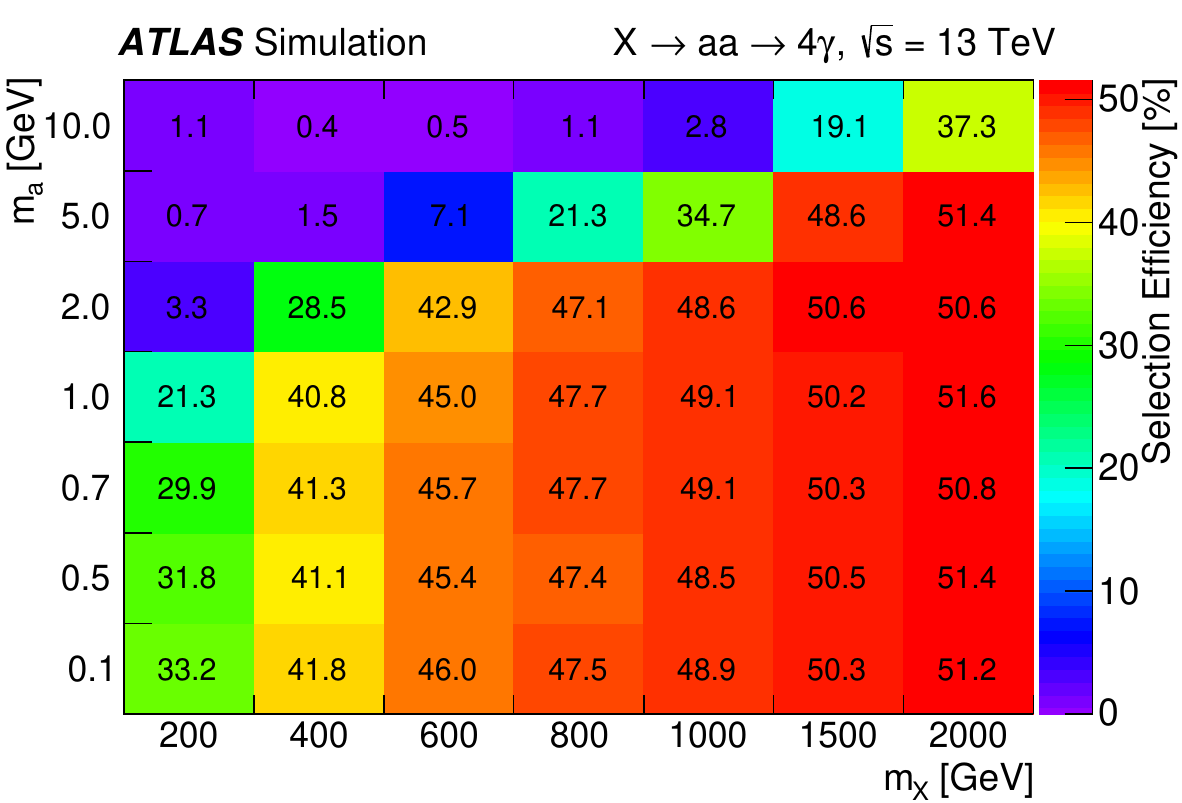}} \hfill
	\subfloat[]{\includegraphics[width=0.48\textwidth]{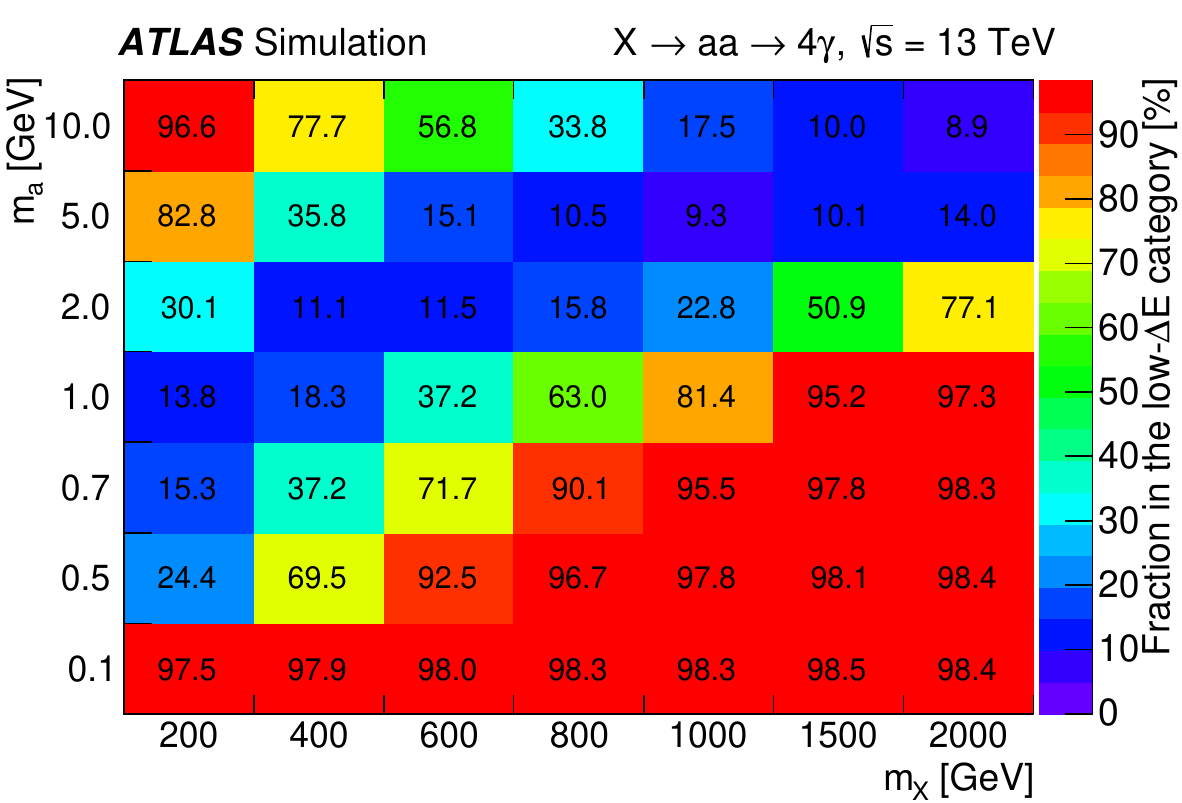}}
	\caption{
		(a) Total selection efficiency $\varepsilon$ (including kinematic acceptance and excluding categorization by $\Delta E$) as a function of $ \ma $ and $ \mX $ for the decay $X\to aa\to4\gamma$. 
		(b) The fraction $f$ of events in the low-$\Delta E$ category.
	}
	\label{fig:TotEff} 
\end{figure}

The overall efficiency, $\varepsilon$, of selecting signal events after applying all criteria, including kinematic acceptance and excluding the categorization by $\Delta E$, is shown in Figure~\ref{fig:TotEff}~(a), and
the fraction, $f$, of signal events that appear in the low-$\Delta E$ category is shown in Figure~\ref{fig:TotEff}~(b), both as a function of \ma\ and \mX\ for the decay $X\to aa\to4\gamma$.
The selection efficiency is low for small values of \mX\ and large values of \ma, and almost all events are in the low-$\Delta E$ category for large values of \mX\ and small values of \ma. For smaller $ m_{a} $ and larger $m_X$, $ f $ increases because of the small angular spread of photons inside the photon-jet, which leads to a calorimeter signature similar to that of a single photon.
Additionally, for larger $ m_{a} $ and smaller $m_X $, $ f $ also increases because individual photons are reconstructed separately due to the large angular separation, resulting in events containing more than two reconstructed photons, each of which more resembles a single photon.
The results for the decay $ X\to aa\to 6\pi^0 $ are similar to those of the decay $ X\to aa\to 4\gamma $.

Table~\ref{tab:datacutflow} displays the number of events in data that satisfy each selection criterion. 
The fraction of events with both of the two leading reconstructed photons found in $ |\eta|<1.37$ (i.e. the barrel section) is 59\% (63\%) for the low-$ \Delta E $ (high-$ \Delta E $) category.

\sisetup{scientific-notation=true, round-mode=figures, round-precision=2}
\begin{table}[h!]
	\begin{center}
		\caption{Number of events in collision data that satisfy the successive selection criteria, as well as the cumulative and relative fraction of events remaining after applying each criterion. The values in the last two lines of the ``Relative'' column are the fraction of events relative to the ``$ \mgg > 175~\GeV $'' line. The values in the ``Preselection'' line include the offline loose photon selection, $ \ET>25~\GeV $, $ |\eta|<2.37 $, excluding the transition region between the barrel and endcap calorimeters, and the matching of the reconstructed photon to the photon trigger object applied to the two leading reconstructed photons. 
	The label ``Relative $ \ET $'' denotes the requirements on $ \ET / \mgg$ for the reconstructed photons, described in Section~\ref{sec:ObjAndEvtSel_BaselineEventSelection}.
}
		\begin{tabular}{lccc}
			\hline
			                                & $N_{\text{observed}}$ &   \multicolumn{2}{c}{Fraction of events}    \\
			      \cmidrule(lr){3-4}        &                       &      Cumulative      &       Relative       \\ \hline
			     All triggered events       &  $ 6.4\times10^9 $    &         ---          &         ---          \\
			         Preselection          &    $ 3.1\times10^7 $     & $4.8 \times 10^{-3}$ & $4.8 \times 10^{-3}$ \\
			Loose$^\prime$ photon selection &    $ 1.7\times10^7 $     & $2.6 \times 10^{-3}$ & $5.3 \times 10^{-1}$ \\
			       Photon isolation         &     $ 2.2\times10^6 $     & $3.4\times 10^{-4}$  & $1.3\times 10^{-1}$  \\
			       Relative $\ET$         &    $ 1.7\times10^6 $     & $2.6\times 10^{-4}$  & $7.7\times 10^{-1}$  \\
			      $ \mgg > 175 $ GeV      &      $ 6.7\times10^4 $    & $1.0\times 10^{-5}$  & $4.0\times 10^{-2}$  \\ \hline
			    Low-$\Delta E$ category     &     $ 5.6\times10^4 $      & $8.8\times 10^{-6}$  & $8.5\times 10^{-1}$  \\
			   High-$\Delta E$ category     &     $ 1.0\times10^4 $      & $1.6\times 10^{-6}$  & $1.5\times 10^{-1}$  \\ \hline
		\end{tabular}
		\label{tab:datacutflow}
	\end{center}
\end{table}

\section{Signal and background modeling}  \label{sec:Bkgds}
The reconstructed signal mass shape is modeled with a double-sided Crystal Ball (DSCB) function.
The backgrounds are determined by fitting functions to the observed mass spectra of two reconstructed photons, \mgg.

\subsection{Signal modeling}
\label{sec:SignalModelling}

The DSCB function has been shown to be effective in modeling new-particle resonances expected to have a Gaussian core  surrounded by asymmetric and non-Gaussian low- and high-mass tails, and is described in detail elsewhere~\cite{Aaboud:2016tru}.
In this analysis the DSCB is a function of the mass of two reconstructed photons (photon-jets in simulated signal samples), with parameters to account for the peak position and width of the Gaussian part, as well as for the upper and lower tails where the resonance shape meets the smoothly falling two-photon mass background.
For the benchmark signal scenario investigated here, since the reconstructed photons are photon-jets (e.g., $ a\to\gamma\gamma $ and $ a\to 3\pi^0\to6\gamma $), the reconstructed $ \mgg $ corresponds to the mass of two $ a $ particles, i.e., the mass of the parent particle, $X$.

For the benchmark signal scenario, for very small values of $m_a / m_X$, the behavior of the DSCB as a function of the mass of two photon-jets is nearly identical to that of the BSM process $X \to 2\gamma$.
The position of the fitted peak of the DSCB is slightly lower than the mass input to the generator.
With the NWA approach, the width of the Gaussian core $ \sigma_\text{CB} $ is dominated by detector resolution, and it increases linearly with $ m_X $, from 2~\GeV\ for $ m_{X} = 200~\GeV $ to 14~\GeV\ for $ m_X = 2~\TeV $.   
For larger values of $m_a / m_X$, the wider opening angle between the photons inside a photon-jet leads to a greater fraction of the energy of the shower leaking out of the window defined by the cells of the EM calorimeter to collect energy for the reconstruction of photons, leading to a further increase in the mass shift and width of the DSCB\@.
For instance, for $ m_X=600~\GeV $ and $ m_a=5~\GeV $, the width is $ \sigma_\text{CB}=8~\GeV $ for the process $ X\to aa\to 4\gamma $, and $ \sigma_\text{CB}=9~\GeV $ for the process $ X\to aa\to 6\pi^0 $.
For a given $ m_X $ and $ m_a $, the same signal mass shape modeling results are used for the analysis of the two orthogonal event categories (the low-$ \Delta E $ category and the high-$ \Delta E $ category), since only a small dependence of the signal mass distributions on $ \Delta E $ is observed.

To validate the mass shape modeling results, injection tests are performed, where a fixed number of signal events are inserted into a pseudo-dataset reproducing a background-only $ \mgg $ spectrum of one of the two event categories, and the number of events inserted is then compared with the number determined by fitting the DSCB\@.
The pseudo-datasets are generated from background probability density functions [represented by Eq.~(\ref{eq:backgroundPdf}), described in Section~\ref{sec:Bkg_Modelling}] with the parameters determined from a fit to the observed $ \mgg $ spectra in collision data.
For each simulated sample of the benchmark scenario, with different values of $m_a$ and $m_X$, separate tests are performed for an increasing number of injected signal events.
The average of the number of events determined from the fit to multiple pseudo-datasets and the number inserted should be identical in an ideal case, and the difference between these two numbers is taken as a systematic uncertainty in the signal mass shape modeling.

The fraction, $f$, of signal events that appear in the low-$\Delta E$ category is parametrized as a function of the mass parameters $m_X$ and $m_a$ of the benchmark signal scenario, to have a continuous model for all the masses considered in the results.
The values of $f$ are taken from simulation and a third-order spline interpolation is performed as a function of $m_a / m_X$.

Similarly, the total signal selection efficiency, $\varepsilon$, is calculated from the individual signal mass points generated, and is parametrized as a function of $m_X$ and $m_a$.
This serves as an input to the calculation of the cross section times branching ratios for the benchmark signal scenario.

The modeling of signal mass shape, $ f $, and  $ \varepsilon $ as functions of $ (m_X, m_a) $ is performed separately for the two different final states of the benchmark signal scenario, $ X\to aa\to4\gamma $ and $ X\to aa\to 6\pi^0 $. 
In general, the results are similar for the two decay scenarios. 
The main distinction is in the different trend in $ f $ with respect to $ m_X $ and $ m_a $, especially the threshold in $ m_a/m_X $ at which the values of $ f $ transition from $ f>0.5 $ to $ f<0.5 $.  This threshold is found to be at $ m_a/m_X\simeq 0.0015 $ for $ X\to aa\to 4\gamma $, and at $ m_a/m_X\simeq 0.0020 $ for $ X\to aa\to 6\pi^0 $.

\subsection{Background modeling} \label{sec:Bkg_Modelling}

The backgrounds in this search mainly consist of the SM production of events containing either two prompt photons; one prompt photon and one hadronic jet; or two hadronic jets.
Prompt photons are defined as photons not originating from hadron decays.
Hadronic jets can be misreconstructed as a photon.
The three background components are denoted \yy, \yj\ or \jy, and \jj, respectively, with the first symbol indicating the one with a higher value of $ \ET $.
The $ \mgg $ distribution of the sum of these background components is described by an analytic function, separately for each of the two $ \Delta E $ categories.
The parameters of the two analytic functions are determined from fits to the $ \mgg $ distributions in the analysis signal region of collision data from a lower edge of $ \mgg = 175~\GeV $.

Based on simulated samples, the contribution from Drell-Yan processes, where the two isolated electrons are misreconstructed as photons, is expected to be at the sub-percent level in the analysis signal region.
The shape of the $ \mgg $ distribution of the Drell-Yan contribution in the mass range $ \mgg>175~\GeV $ is expected to be similar to that of the $ \yy $ component, and it is therefore absorbed into the analytic function fit for the continuum background components.

The choice of the functional form describing the background distribution is based on studies of background templates.
A variety of functional forms are considered for the background parametrization to achieve a good compromise between limiting the size of a potential bias toward the identification of a signal when none is present (the \textit{spurious signal}) and retaining good statistical power.
The size of the spurious signal for a given functional form is estimated by performing a maximum-likelihood fit to the background templates using the sum of signal and background parametrizations. 

To determine the overall shape of the background mass spectra, background templates are determined using both the simulation and collision data, separately for each of the two $ \Delta E $ categories.
A simulated sample of prompt diphoton events is used to model the shape of the contribution from \yy\ events.
Subsets of collision data that are similar but orthogonal to the signal region are used to determine the shapes of the \yj, \jy\ and \jj\ components, where the subleading reconstructed photon, leading reconstructed photon, or both reconstructed photons, respectively, are required to fail the default isolation criterion but satisfy a looser one. 
This looser criterion is defined by loosening the requirement for the calorimeter isolation variable to $ \ET^{\text{iso}} < 7~\GeV $.
The resulting samples of $ \yy, \yj, \jy $, and $ \jj $ are summed to derive the background templates, scaled with the background composition fractions determined from the matrix method described below.

The background composition of a given mass spectrum of two reconstructed photons is estimated using a matrix method~\cite{Aad:2011mh}, where events are categorized into four subsets by whether both, only the leading, only the subleading, or neither of the two leading reconstructed photons satisfy the calorimeter isolation requirement.
The method relies on external estimates of the efficiency for prompt photons satisfying calorimeter isolation and the rate at which hadronic jets can mimic a photon satisfying calorimeter isolation (the ``fake rate'').
Photon isolation efficiency is estimated with simulated samples of prompt photons.
The isolation variables of photons in simulated samples are adjusted by applying correction factors obtained from small differences observed between photon-enriched control samples of collision data and simulation.
An uncertainty is assessed for the photon isolation efficiency by comparing the nominal efficiencies with those derived without applying the corrections to the isolation variable in simulated samples.
Fake rates are determined using subsets of collision data with selection criteria imposed so that they are similar but orthogonal to the analysis signal region (``control regions'').
These control regions are defined by requiring reconstructed photons to fail the baseline loose$^\prime$ photon selection but satisfy another, looser photon selection.
This looser photon selection, with respect to the loose$^\prime$ selection, is defined by removing requirements on two additional shower shape variables that quantify the lateral shower development in the EM calorimeter second layer ($ w_{\eta 2} $ and $ R_{\phi} $, described in Table 1 of Ref.~\cite{Aaboud:2016yuq}).
A difference of approximately $ 1~\GeV $ is found between the isolation energy spectra in the signal and control regions.
This is accounted for by shifting the threshold of the isolation selection criteria by $\pm$ 1~\GeV, determining the resulting change in the calculated fake rates, and assigning the difference as a systematic uncertainty in these values.

An additional uncertainty is assessed by altering the definition of the control regions.
To accomplish this, a looser photon selection, with respect to the loose$^\prime$ selection, is defined by removing the requirement on one shower shape variable ($ w_{\eta 2} $) instead of two and comparing the difference between the resulting fake rates.

\begin{figure}[h!]
	\centering
	\subfloat[]{\includegraphics[width=0.49\textwidth]{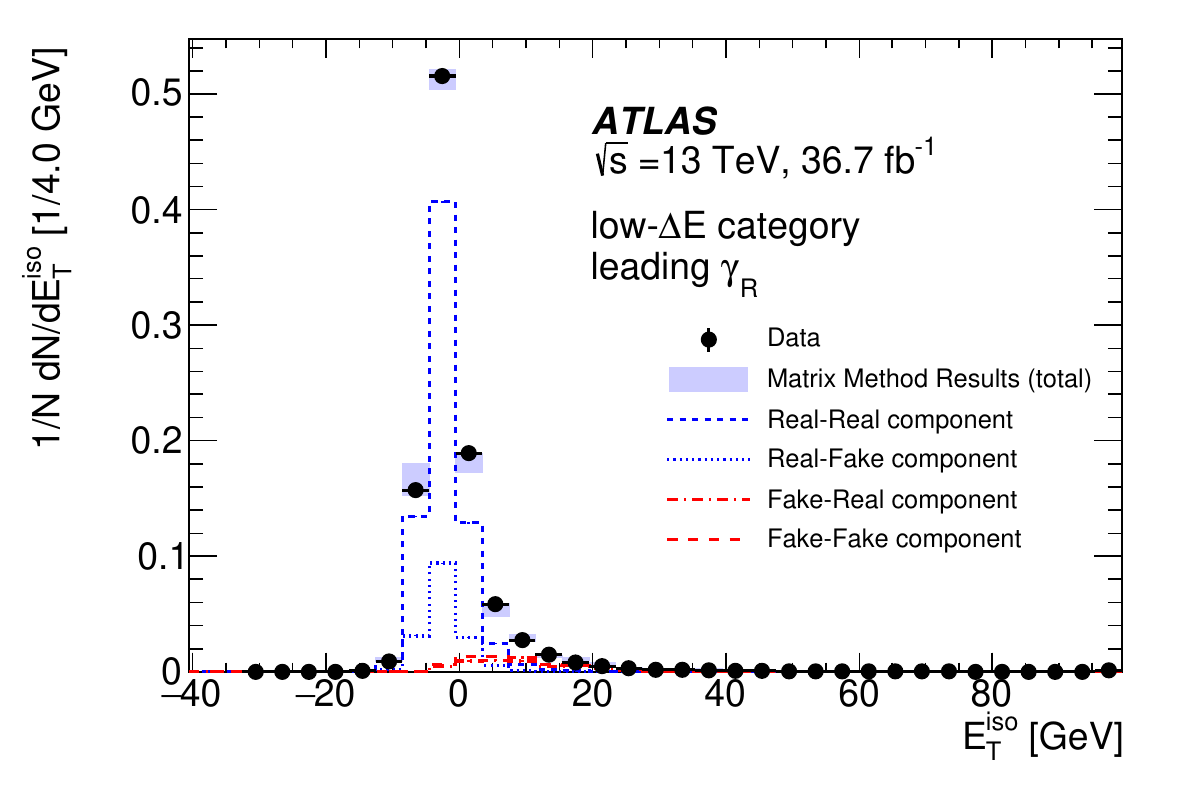}}
	\subfloat[]{\includegraphics[width=0.49\textwidth]{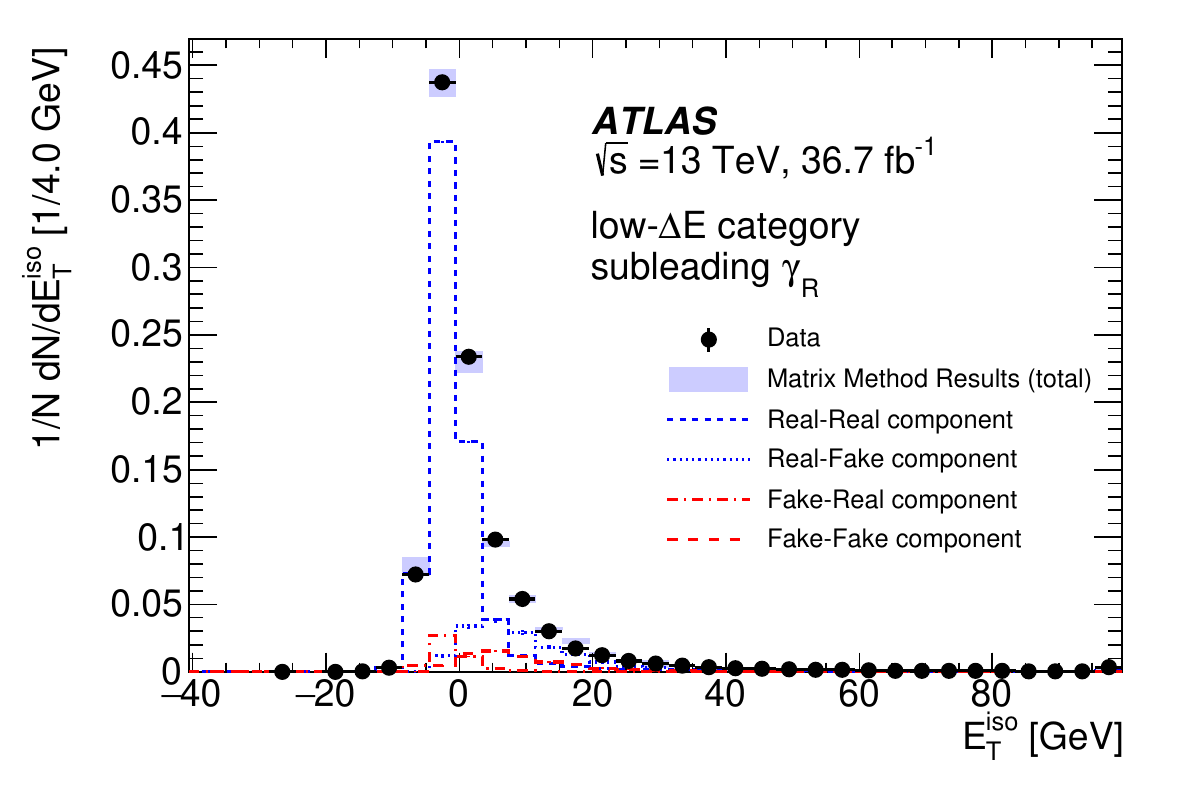}} \\
	\subfloat[]{\includegraphics[width=0.49\textwidth]{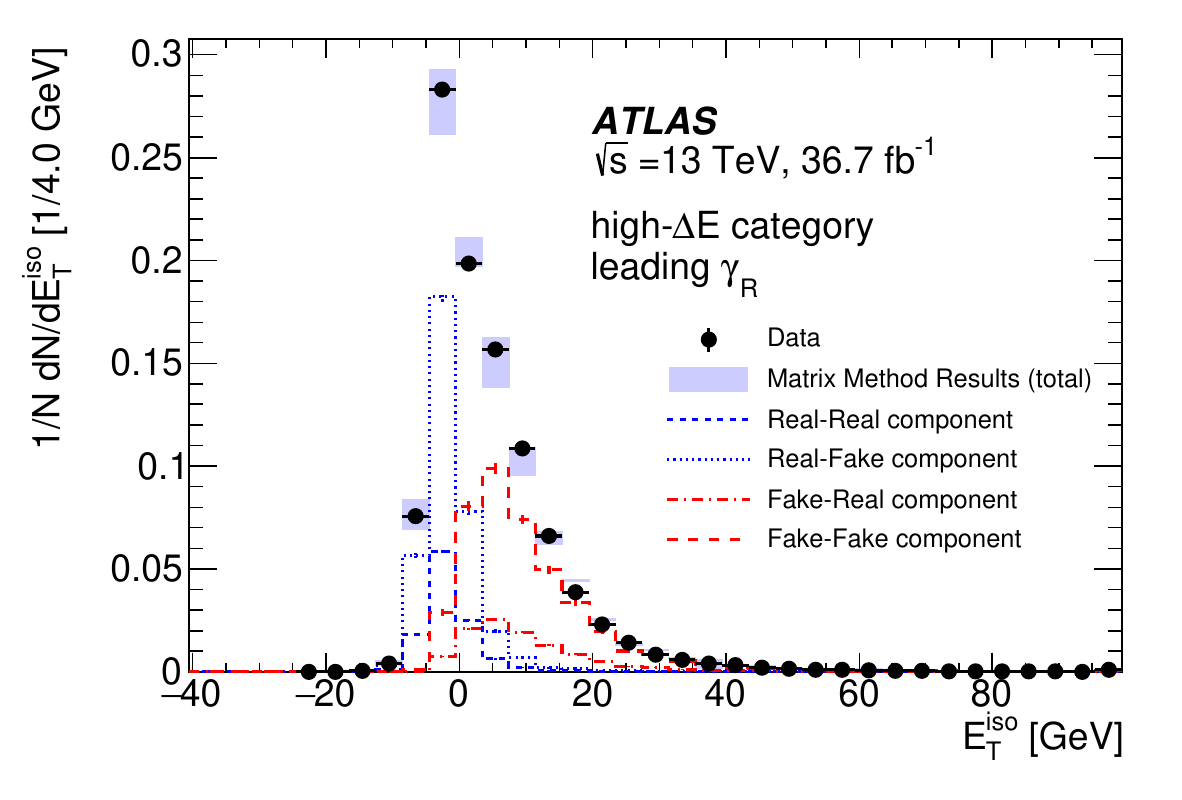}}
	\subfloat[]{\includegraphics[width=0.49\textwidth]{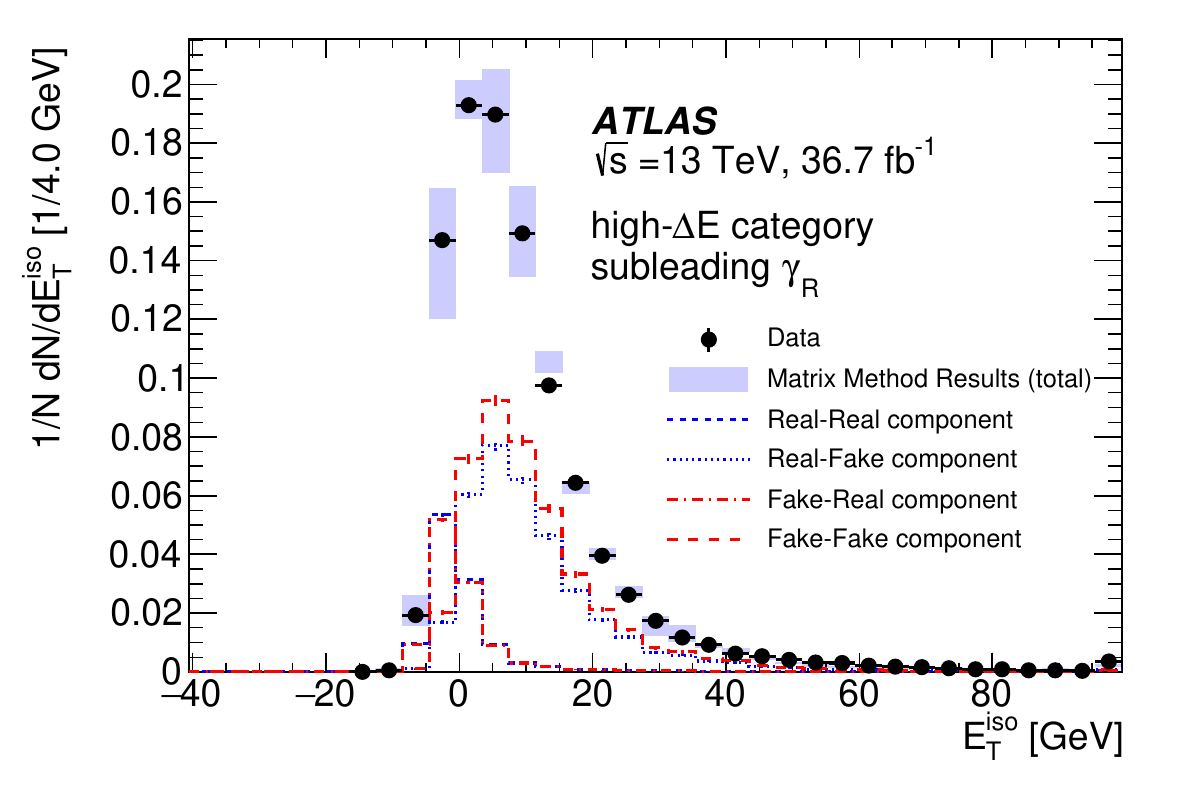}} \\
	\caption{
		Comparison of the observed $ \ET^{\text{iso}} $ spectra and the expected spectra based on the background composition measurement results.
		The modeled spectra of $\gamma \gamma$ (dashed), $\gamma j$ (dotted), $j \gamma$ (dot-dashed), and $jj$ (long-dashed) components are added using the background composition measured with the matrix method.
		The results are compared for each of the two $ \Delta E $ categories where (a) shows the leading reconstructed photon in the low-$ \Delta E $ category, (b) the subleading reconstructed photon in the low-$ \Delta E $ category, (c) the leading reconstructed photon in the high-$ \Delta E $ category, and (d) the subleading reconstructed photon in the high-$ \Delta E $ category. }
	\label{fig:purity_caloisolationDistribution_matrixmethodResults}
\end{figure}

The resulting background compositions are shown in Table~\ref{tab:background_composition_summary}.
Good agreement is seen between the observed isolation spectrum and the expected spectrum based on the matrix method results, within uncertainties, as shown in Figure~\ref{fig:purity_caloisolationDistribution_matrixmethodResults}.

\renewcommand{\arraystretch}{1.2}
\begin{table}[h!]
	\begin{center}
		\caption{ Summary of the measured background compositions for the two categories.}
		\begin{tabular}{|c|c|c|}
			\hline
			                &    Low-$\Delta E$ category    &  High-$\Delta E$ category  \\ \hline
			$\gamma \gamma$ &  $ 0.930^{+0.027}_{-0.031} $  &      $0.48 \pm 0.16 $      \\
			  $\gamma j$    & $ 0.051 ^ {+0.021}_{-0.018} $ &  $ 0.32 ^{+0.08}_{-0.09}$  \\
			  $j \gamma$    &  $0.014 ^{+0.004}_{-0.005}$   & $0.108 ^{+0.001}_{-0.016}$ \\
			     $j j$      &  $ 0.005^{+0.006}_{-0.003} $  &  $0.09 ^{+0.09}_{-0.05}$   \\ \hline
		\end{tabular}
		\label{tab:background_composition_summary}
	\end{center}
\end{table}
\renewcommand{\arraystretch}{1.}

The background templates are derived with the summation of the $ \yy, \yj, \jy $ and $ \jj $ components scaled by the background composition fractions, separately for each of the two $ \Delta E $ categories, as described above.
The resulting background templates are presented in Figure~\ref{fig:backgroundmodelingtemplate}.

To evaluate the size of the spurious signal, a test is performed using these background templates and the signal modeling described in Section~\ref{sec:SignalModelling}.
The background templates are normalized to the integrated luminosity for this search, 36.7~\ifb.
A family of functions, adapted from those used by searches for new physics signatures in dijet final states~\cite{Aaltonen:2008dn}, is chosen to describe the shape of the $ \mgg $ distribution:
\begin{equation}\label{eq:backgroundPdf}
g_{(k)}\left(x;a, \left\{b_j\right\}_{j=0,k} \right)= N\left(1-x^\frac{1}{2}\right)^ax^{\sum_{j=0}^k b_j(\log x)^j}
\end{equation}
The variable $x$ is defined as $x=\mgg / \sqrt{s} $. The parameters $ a $ and $ b_j $ are free parameters and $ N $ is the normalization factor.
The spurious signal tests are then performed using a maximum-likelihood fit of the sum of the signal and background parametrizations to each of the two background templates.
The spurious signal is allowed to be negative as well as positive.
The final functional form used to model the background when performing the search for resonances is one where the estimated spurious signal is required to be smaller than 30\% of the statistical uncertainty in the fitted signal yield across the full mass spectrum.
The cutoff of $ 30\% $ is chosen to ensure that the contribution of this systematic uncertainty to the total uncertainty, including all statistical and systematic uncertainties, is subdominant and smaller than 5\%.

The method is validated by checking that similar results are obtained when the test is performed using variations of the background templates, for which the background compositions are shifted within the uncertainties presented in Table~\ref{tab:background_composition_summary}.
When the fraction of the $ \gamma\gamma $ component is shifted up and those for $ \gamma j $, $ j\gamma $, $ jj $ are shifted down, or vice versa, the size of the resulting spurious signals are consistent within the statistical uncertainty of the background templates.

The resulting functional form used for the background mass spectrum evaluation of the two categories is shown in Eq.~(\ref{eq:backgroundPdf}) with $ k=1 $.
Figure~\ref{fig:backgroundmodelingtemplate} shows the level of agreement between this functional form and the background templates.
The resulting background model and its associated systematic uncertainties are used when searching for resonances in the mass spectra of the signal region.

\begin{figure}
	\centering
	\subfloat{\includegraphics[width=0.5\textwidth]{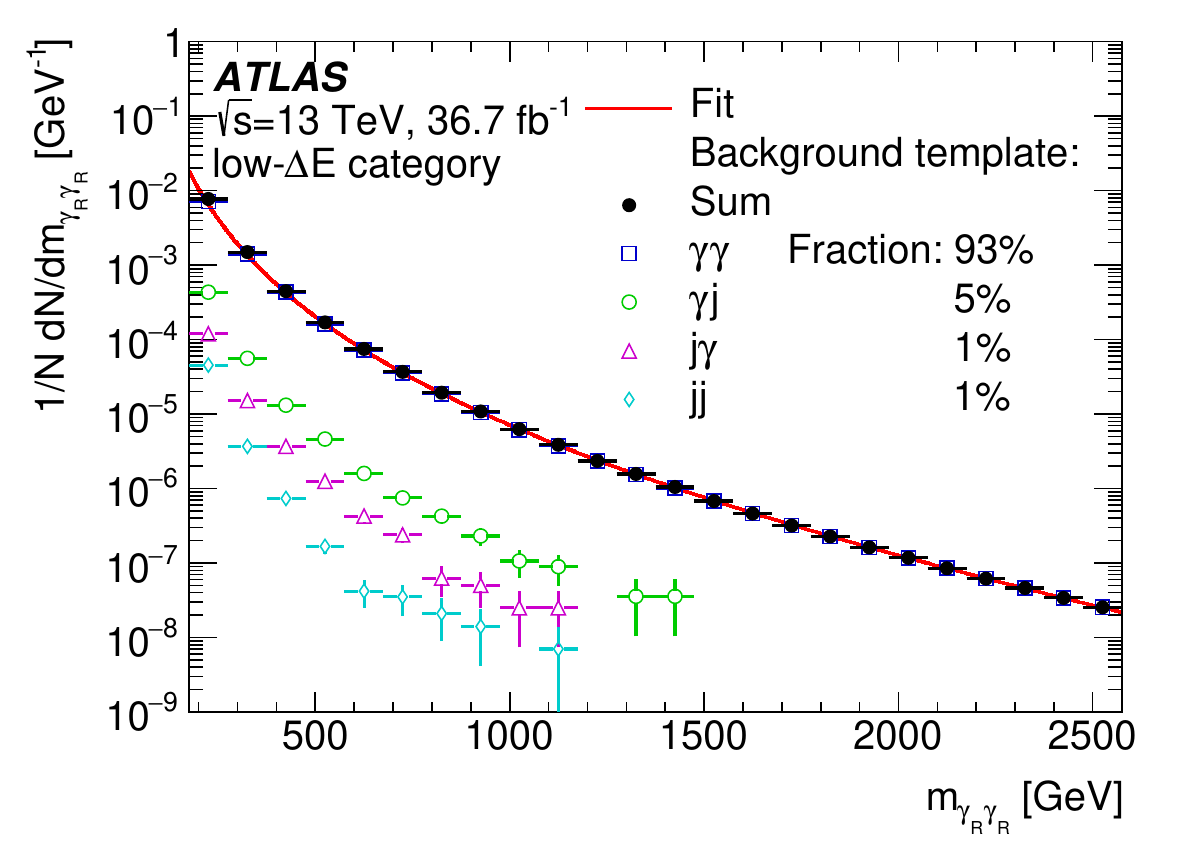} }
	\subfloat{\includegraphics[width=0.5\textwidth]{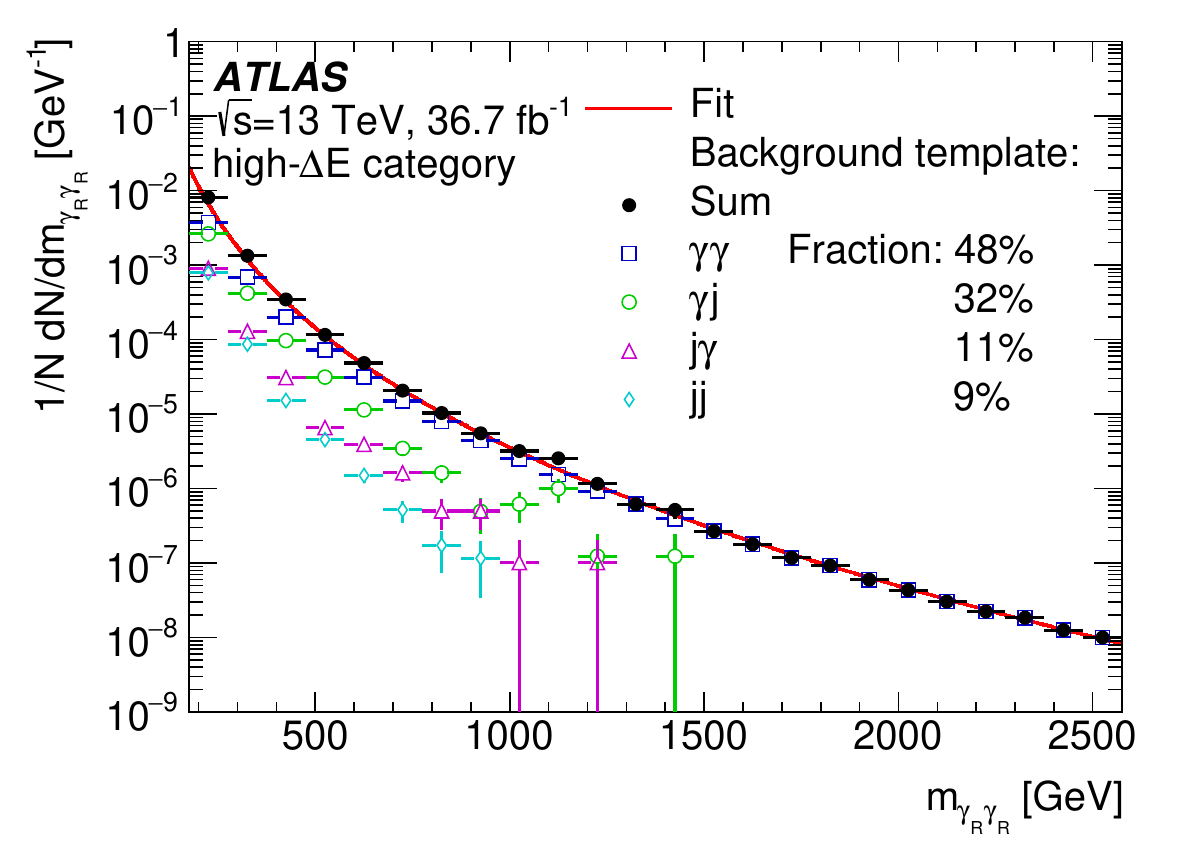} }
	\caption{ Background templates used for the spurious signal test.
                  The sum of the background components for each of the two $ \Delta E $ categories, and the breakdown into components ($ \gamma\gamma $, $ \gamma j $, $ j\gamma $, and $ jj $) are shown.
                  The unbinned likelihood fit with the chosen functional form (shown in Eq.~(\ref{eq:backgroundPdf}) with $ k=1 $) is superimposed.
                  The expected background compositions, which are measured inclusively for events in $ \mgg > 175~\GeV $, are shown on the figures. }
	\label{fig:backgroundmodelingtemplate}
\end{figure}

\section{Systematic uncertainties} \label{sec:SysUnc}
Several sources of systematic uncertainties that affect the determination of the signal yield are taken into account.
In most cases, systematic uncertainties are smaller than statistical errors.

The uncertainty in the combined 2015+2016 integrated luminosity is 2.1\%. It is derived, following a methodology similar to that detailed in Ref.~\cite{Aaboud:2016hhf}, from a calibration of the luminosity scale using $x$--$y$ beam-separation scans performed in August 2015 and May 2016.

The impact of the photon energy resolution on signal modeling is evaluated.
It mainly affects the mass shape width, $ \sigma_\text{CB} $, of the Crystal Ball function used to model the signal mass shape.
The photon energy resolution is adjusted by one standard deviation from the nominal value in both positive and negative directions, and the resulting change in the fitted signal width is determined.
The relative difference in the fitted value of $ \sigma_\text{CB} $ ranges from as small as a few percent to as large as 37\%, increasing with larger $ m_X $ and dependent slightly on $ m_a $, and is taken as a systematic uncertainty.

Systematic uncertainties in the extracted signal yield due to signal mass shape modeling are evaluated via injection tests, described in Section~\ref{sec:SignalModelling}.
The final fitted values of the number of signal events deviate from the injected values by less than 1\% almost everywhere, rising to a maximum of 5\% for some signal mass values at the edge of the analysis search region of $ m_a=0.1~\GeV $ for the high-$ \Delta E $ category and $ m_a =0.01\times m_X$ for the low-$ \Delta E $ category.
This is taken as the estimate of the systematic uncertainty in the signal yield.

Uncertainties in the modeling of the category fraction, $ f $, are evaluated by an envelope to cover the deviations of the values of $ f $ from simulation and the parametrization.
The absolute value of the change in $ f $ varies as a function of $ m_a/m_X $, from 3\% at $ m_a/m_X=0 $, increasing to 12\%--14\% at around $ m_a/m_X=0.002 $, and decreasing to 6\%--10\% at $ 0.002 <m_a/m_X<0.01 $.
This is taken as the estimate of the systematic uncertainty in $f$.

Other systematic uncertainties in the extracted signal yield and the migration of signal events between the two orthogonal $ \Delta E $ categories are evaluated by comparisons between nominal and systematically varied versions of various experimental uncertainty sources, such as the photon energy scale and resolution, isolation selection efficiency, shower shape modeling, and pileup.
The systematic uncertainties due to the photon energy scale and resolution are adapted from results determined during LHC Run 1~\cite{PERF-2013-05}, with minor updates derived from data-driven corrections determined using Run 2 data.
Uncertainties related to the loose$^\prime$ photon identification scheme are evaluated with the systematic variations for the shower shape modeling, without the correction factors applied to simulation derived from small differences observed between photon-enriched control samples of collision data and simulation~\cite{ATL-PHYS-PUB-2016-014}.
The uncertainty in the photon calorimeter isolation efficiency is calculated from changes due to applying and not applying corrections derived from small differences observed between photon-enriched control samples of collision data and simulation.
The uncertainties of the efficiency correction factors using photon-enriched control samples of collision data are used to derive the uncertainty in the photon track isolation efficiency.
The pileup uncertainty is taken into account by propagating it through the event
selection.
The uncertainties in $\varepsilon$ and  $f$ due to these sources for the mass regions considered for the benchmark signal scenario are calculated.
The uncertainties are less than 1\% in almost all cases, rising to $\sim$4\% for some isolation and shower shape uncertainties for larger values of $m_a / m_X$ at the edge of the analysis sensitivity.  

Additional systematic uncertainties in the loose diphoton trigger efficiency are not assessed.
The $\ET$ requirements for reconstructed photons are much larger than the value at which the diphoton trigger utilized becomes nearly 100\% efficient, and any additional uncertainties in signal efficiency due to mismodeling of the trigger-level shower shape variables are accounted for when calculating uncertainties in offline loose$^\prime$ identification, because the loose photon identification definitions at the trigger and offline levels are strongly correlated.

The uncertainty in the signal kinematic acceptance, which is included in the definition of the total signal selection efficiency, is evaluated for the choice of PDF set used for the simulation of the signal samples. It is less than 1\% in most cases, rising to $\sim4\%$ for large $ m_X $ around $ m_X\sim 2~\TeV $.

The systematic uncertainties related to the evaluation of the background mass spectrum are determined from the spurious signal method, described in Section~\ref{sec:Bkg_Modelling}.
The  spurious signal as a function of $ m_X $ and $ m_a $ is parametrized so that the modeling between mass points is continuous.
This parametrization is performed in such a way that it can slightly overestimate the size of spurious signals, especially at the lower end of the $ m_X $ range, $ m_X = 200~\GeV $.
The size of the parametrized spurious signal decreases for larger $ m_X $ and depends slightly on the $ m_a $ value, ranging from 85 to $6\times10^{-3}$ events for the low-$\Delta E$ category, and from 32 to $1\times10^{-2}$ events for the high-$\Delta E$ category.

The systematic uncertainties are generally smaller than the statistical errors, with the systematic uncertainty in the background evaluated from the spurious signal being the largest contribution.
This is because the parametrization of the size of spurious signals slightly overestimates the values at the lower end of the $ m_X $ range, as described above.
The impact of the systematic uncertainties on the expected limit decreases with the resonance mass $m_X$ from 51\% at most for $m_X=200~\GeV$ to 5\% at most for $m_X > 800~\GeV$.
The impact of the systematic uncertainties on the signal yield obtained from the fit is summarized in Table~\ref{tab:impact}.

\newcommand{\ImpactAMx}{200}
\newcommand{\ImpactAMa}{0.1}
\newcommand{\ImpactAStat}{66}
\newcommand{\ImpactAA}{2}
\newcommand{\ImpactAB}{74}
\newcommand{\ImpactAC}{6}
\newcommand{\ImpactAD}{1}
\newcommand{\ImpactAE}{7}
\newcommand{\ImpactAF}{1}
\newcommand{\ImpactAG}{1}
\newcommand{\ImpactAH}{0}
\newcommand{\ImpactAI}{7}
\newcommand{\ImpactAJ}{3}
\newcommand{\ImpactAK}{0}
\newcommand{\ImpactAL}{3}
\newcommand{\ImpactAM}{1}
\newcommand{\ImpactAN}{2}
\newcommand{\ImpactBMx}{200}
\newcommand{\ImpactBMa}{0.3}
\newcommand{\ImpactBStat}{72}
\newcommand{\ImpactBA}{1}
\newcommand{\ImpactBB}{37}
\newcommand{\ImpactBC}{67}
\newcommand{\ImpactBD}{0}
\newcommand{\ImpactBE}{2}
\newcommand{\ImpactBF}{0}
\newcommand{\ImpactBG}{0}
\newcommand{\ImpactBH}{0}
\newcommand{\ImpactBI}{19}
\newcommand{\ImpactBJ}{2}
\newcommand{\ImpactBK}{0}
\newcommand{\ImpactBL}{1}
\newcommand{\ImpactBM}{0}
\newcommand{\ImpactBN}{1}
\newcommand{\ImpactCMx}{200}
\newcommand{\ImpactCMa}{2}
\newcommand{\ImpactCStat}{86}
\newcommand{\ImpactCA}{1}
\newcommand{\ImpactCB}{9}
\newcommand{\ImpactCC}{55}
\newcommand{\ImpactCD}{1}
\newcommand{\ImpactCE}{5}
\newcommand{\ImpactCF}{1}
\newcommand{\ImpactCG}{2}
\newcommand{\ImpactCH}{0}
\newcommand{\ImpactCI}{9}
\newcommand{\ImpactCJ}{1}
\newcommand{\ImpactCK}{1}
\newcommand{\ImpactCL}{0}
\newcommand{\ImpactCM}{1}
\newcommand{\ImpactCN}{2}
\newcommand{\ImpactDMx}{600}
\newcommand{\ImpactDMa}{0.1}
\newcommand{\ImpactDStat}{99}
\newcommand{\ImpactDA}{2}
\newcommand{\ImpactDB}{13}
\newcommand{\ImpactDC}{2}
\newcommand{\ImpactDD}{4}
\newcommand{\ImpactDE}{13}
\newcommand{\ImpactDF}{3}
\newcommand{\ImpactDG}{2}
\newcommand{\ImpactDH}{4}
\newcommand{\ImpactDI}{3}
\newcommand{\ImpactDJ}{3}
\newcommand{\ImpactDK}{3}
\newcommand{\ImpactDL}{5}
\newcommand{\ImpactDM}{3}
\newcommand{\ImpactDN}{2}
\newcommand{\ImpactEMx}{600}
\newcommand{\ImpactEMa}{0.9}
\newcommand{\ImpactEStat}{94}
\newcommand{\ImpactEA}{3}
\newcommand{\ImpactEB}{5}
\newcommand{\ImpactEC}{24}
\newcommand{\ImpactED}{4}
\newcommand{\ImpactEE}{12}
\newcommand{\ImpactEF}{4}
\newcommand{\ImpactEG}{3}
\newcommand{\ImpactEH}{4}
\newcommand{\ImpactEI}{25}
\newcommand{\ImpactEJ}{4}
\newcommand{\ImpactEK}{3}
\newcommand{\ImpactEL}{4}
\newcommand{\ImpactEM}{3}
\newcommand{\ImpactEN}{3}
\newcommand{\ImpactFMx}{600}
\newcommand{\ImpactFMa}{5}
\newcommand{\ImpactFStat}{98}
\newcommand{\ImpactFA}{3}
\newcommand{\ImpactFB}{3}
\newcommand{\ImpactFC}{22}
\newcommand{\ImpactFD}{2}
\newcommand{\ImpactFE}{1}
\newcommand{\ImpactFF}{3}
\newcommand{\ImpactFG}{3}
\newcommand{\ImpactFH}{3}
\newcommand{\ImpactFI}{7}
\newcommand{\ImpactFJ}{4}
\newcommand{\ImpactFK}{2}
\newcommand{\ImpactFL}{3}
\newcommand{\ImpactFM}{3}
\newcommand{\ImpactFN}{2}

\begin{table}[h!]
	\begin{center}
		\caption{ 
	Breakdown of the relative contributions to the total uncertainty in the signal yield obtained from the fit. For each source of uncertainty $ \sigma_\text{source} $, the fraction $ \sigma_\text{source} / \sigma_\text{total} $ is presented, where $ \sigma_\text{total} $ is the total uncertainty that includes the statistical uncertainty. The sum in quadrature of the individual components differs from 100\% due to small correlations between the components.
	The values here are for the signal process $ X\to aa\to 4\gamma $.
	The mass points $ (m_X,m_a)=(200~\GeV,0.3~\GeV),(600~\GeV,0.9~\GeV) $ correspond to those values for which the systematic uncertainty of the category fraction $ f $ is the highest.
	Similar results are found for the decay $ X\to aa\to 6\pi^0 $.
	}
\begin{tabular}{lcccccc}
	\hline
	         $ m_X~[\GeV],m_a~[\GeV] $          &  $ (\ImpactAMx ,\ImpactAMa) $  &  $ (\ImpactBMx ,\ImpactBMa) $  &  $ (\ImpactCMx ,\ImpactCMa) $  &  $ (\ImpactDMx ,\ImpactDMa) $  &  $ (\ImpactEMx, \ImpactEMa) $  &  $ (\ImpactFMx, \ImpactFMa) $  \\ \hline
	                                            &                                                                        \multicolumn{6}{c}{ $ \sigma_\text{source}/\sigma_\text{total} $ }                                                                         \\
	\cmidrule(lr){2-7}  
	Statistical & $ \ImpactAStat    \%         $ & $ \ImpactBStat    \%         $ & $ \ImpactCStat      \%       $ & $ \ImpactDStat    \%         $ & $ \ImpactEStat     \%        $ & $ \ImpactFStat     \%        $ \\
	    Spurious signal (low-$ \Delta E $)      &  $ \ImpactAB    \%         $   &  $ \ImpactBB    \%         $   &  $ \ImpactCB      \%       $   &  $ \ImpactDB    \%         $   &  $ \ImpactEB     \%        $   &  $ \ImpactFB     \%        $   \\
	    Spurious signal (high-$ \Delta E $)     &  $ \ImpactAC    \%         $   &  $ \ImpactBC    \%         $   &  $ \ImpactCC      \%       $   &  $ \ImpactDC    \%         $   &  $ \ImpactEC     \%        $   &  $ \ImpactFC     \%        $   \\
	          Category fraction $ f $           &  $ \ImpactAI    \%         $   &  $ \ImpactBI    \%         $   &  $ \ImpactCI      \%       $   &  $ \ImpactDI    \%         $   &  $ \ImpactEI     \%        $   &  $ \ImpactFI     \%        $   \\
	          Signal mass resolution            &  $ \ImpactAE    \%         $   &  $ \ImpactBE    \%         $   &  $ \ImpactCE      \%       $   &  $ \ImpactDE    \%         $   &  $ \ImpactEE     \%        $   &  $ \ImpactFE     \%        $   \\
	   Signal mass shape (low-$ \Delta E $)     &  $ \ImpactAL    \%         $   &  $ \ImpactBL    \%         $   &  --   &  $ \ImpactDL    \%         $   &  $ \ImpactEL     \%        $   &  $ \ImpactFL     \%        $   \\
	   Signal mass shape (high-$ \Delta E $)    &  --            &  --            &  $ \ImpactCK      \%       $   &  $ \ImpactDK    \%         $   &  $ \ImpactEK     \%        $   &  $ \ImpactFK     \%        $   \\ \hline
\end{tabular} 
\label{tab:impact}
\end{center}
\end{table}

\section{Statistical procedure} \label{sec:Stat}
For a given fixed signal mass hypothesis, a mass spectrum fit including both the background and signal components is performed to the full mass spectrum of $\mgg >$ 175~\GeV, using an unbinned maximum-likelihood approach, simultaneously for the two categories (low-$\Delta E$ and high-$\Delta E$ categories).  
A constraint is placed on the ratio of the two separate normalization factors of the signal component for the two categories, evaluated from the category fraction $f$, which depends on the signal masses $ m_X $ and $ m_a $.
Deviations from the background-only hypothesis are searched for starting from $m_{X}$ = 200~\GeV, and the entire $ \mgg $ range is used for the background component for each hypothesis test.
The $p$-values are calculated with the profile likelihood ratio as the basis for the test statistic and utilizing an asymptotic approximation~\cite{Cowan:2010js}. 

Systematic uncertainties (described in Section~\ref{sec:SysUnc}) are treated as nuisance parameters in the likelihood function, where each is a floating parameter constrained by either a Gaussian function (for spurious signal and uncertainties related to the migration of events between the $\Delta E$ categories) or a log-normal function (for all other uncertainties).
Two nuisance parameters are introduced for the extracted signal yield due to signal mass shape modeling uncertainties, one for each $ \Delta E $ category, and they are multiplied by the signal normalizations of each category.
One nuisance parameter is introduced for the impact of the photon energy resolution on the mass shape width, and it is multiplied by the signal mass shape width $ \sigma_\text{CB} $.
One nuisance parameter is introduced for the modeling of the category fraction, $ f $, which is added to $ f $ to shift its value.
Several nuisance parameters are introduced for experimental uncertainty sources and PDF uncertainty that affect the extracted signal yield, total signal selection efficiency, $ \varepsilon $, and category fraction, $ f $.
Two nuisance parameters are introduced for the spurious signals, one for each $ \Delta E $ category.
For a given signal mass $ (m_X,m_a) $ hypothesis, the spurious signals are given the same $ \mgg $ shape as the signal component, and normalized by the size of the spurious signals. 

The calculation of $p$-values for the background-only hypothesis ($p_0$) is performed for a narrow resonance from $ m_X=200~\GeV $ to $ m_X=2.7~\TeV $, with a scan step of $ 1~\GeV $.
Since the samples for the benchmark signal scenario were simulated for the $ m_X $ values in the range $ 200~\GeV < m_X < 2~\TeV $, the results of signal mass shape modeling, modeling of category fraction $ f $, and systematic uncertainties are extrapolated for the $ m_X $ values in the range $ 2~\TeV<m_X<2.7~\TeV $. 

Expected and observed upper limits, at the 95\% confidence level (C.L.), on the production cross section times the product of branching ratios are calculated as a function of the mass parameters of the benchmark signal scenario, $m_X$ and $m_a$, following the CL$_\text{s}$ modified frequentist prescription~\cite{0954-3899-28-10-313}.
Upper limits are determined separately for the two final states of the benchmark signal scenario where the $a$ particle decays into either a pair of photons or three neutral pions.

The assumptions inherent in the use of the asymptotic approximation are validated by sampling distributions of the test statistic using pseudoexperiments, for a few signal mass points. 
The asymptotic approximation yields median values of the expected upper limits within 5\% of those calculated with a large number of pseudoexperiments for most of the values of $m_{X}$ and $m_{a}$ tested.
Due to the small number of events in data in the region $ \mgg>1~\TeV $ in the high-$ \Delta E $ category, larger deviations are observed for $ m_{X} > 1~\TeV $ and large $ m_a $.
The deviation is smaller than 5\% at $ (m_X,m_a)=(1~\TeV,10~\GeV) $, but the expected upper limits obtained from the asymptotic approximation are smaller than those from pseudoexperiments by 20\% for $ (m_X,m_a)=(1.5~\TeV,10~\GeV) $, and 30\% for $ (m_X,m_a)=(2~\TeV,10~\GeV) $.

\section{Results}  \label{sec:Results}
The observed $\mgg$ spectra in the signal region are shown in Figure~\ref{fig:results_fit}.
The results of the two-dimensional scan of $ p_0 $, equivalently expressed in terms of the local significance---the number of standard deviations away from the mean of a normal distribution---are shown in Figure~\ref{fig:results_significance}.
Two different regimes can be seen in this plot, above and below the threshold at $m_a \sim 0.0015 \times m_X $.
These are a result of the categorization of events based on the $\Delta E$ variable.
For $m_a \lesssim 0.0015 \times m_X $, a larger fraction of signal events is expected in the low-$\Delta E$ category, and for $m_a \gtrsim 0.0015 \times m_X$, a larger fraction of signal events is expected in the high-$\Delta E$ category.
The largest local deviation from the background-only hypothesis is found to be $2.7\sigma$, corresponding to $m_X$ = 729~\GeV\ and $m_a$ = 0.1~\GeV\ for the decay $X\to aa\to4\gamma$. 
The width of the signal mass shape for $ m_X=729~\GeV $ and $ m_a=0.1~\GeV $ is $ 6~\GeV $, and thus this deviation appears as a small area in Figure~\ref{fig:results_significance}.
A small excess of events is also observed centered around $ m_X=1.1~\TeV $ and $ m_a=7~\GeV $, which corresponds to a local deviation of $ 2.2\sigma $.
The observed maximum local deviation is less significant than the median of the largest deviation obtained in background-only pseudoexperiments, calculated in the search region defined by $ m_X $ values from $ 200~\GeV $ to $ 2.7~\TeV $ and $ m_a $ values from $ 0.1~\GeV $ to $ 0.01\times m_X $.
The $ \mgg $ mass distribution is found to be consistent with the background-only hypothesis.

The 95\% C.L. observed and expected upper limits on the cross section for the production via gluon-gluon fusion of a high-mass scalar particle, $X$, with narrow width times the branching ratios into a pair of $a$ particles and the subsequent decay of each $a$ into a pair of photons, $\sigma_{X} \times {\mathcal {B}}(X \rightarrow aa) \times {\mathcal {B}}(a \rightarrow \gamma\gamma)^{2}$, are shown in Figure~\ref{fig:results_limit_superimposed}, separately for different values of $m_{a}$.
The same result is presented in Figure~\ref{fig:results_limit_superimposed_mamX}, with the ratio $m_a/m_X$ shown on the horizontal axis.
This plot illustrates the two features of this search.
First, when the ratio $m_a/m_X$ is larger than a threshold of roughly $0.0015$, more signal events are expected in the high-$\Delta E$ category, which has a significantly better signal-to-background ratio compared with the low-$\Delta E$ category, thus leading to stronger upper limits.
Second, for larger values of $m_a/m_X$, the decrease in the signal selection efficiency leads to weaker upper limits.

The 95\% C.L. observed and expected upper limits on the cross section times product of branching ratios for the decay of the $a$ into three neutral pions, $\sigma_{X} \times {\mathcal {B}}(X \rightarrow aa) \times {\mathcal {B}}(a \rightarrow 3\pi^{0})^{2}$, are shown in Figure~\ref{fig:results_limit_superimposed_sixpion}, separately for different values of $m_{a}$.
This result shows features similar to that shown in Figure~\ref{fig:results_limit_superimposed}, with slight differences arising mainly from the different trend of the category fraction, $f$, with respect to the mass values $m_X$ and $m_a$.

\begin{figure}[h!]
	\centering
	\subfloat{\includegraphics[width=0.5\textwidth]{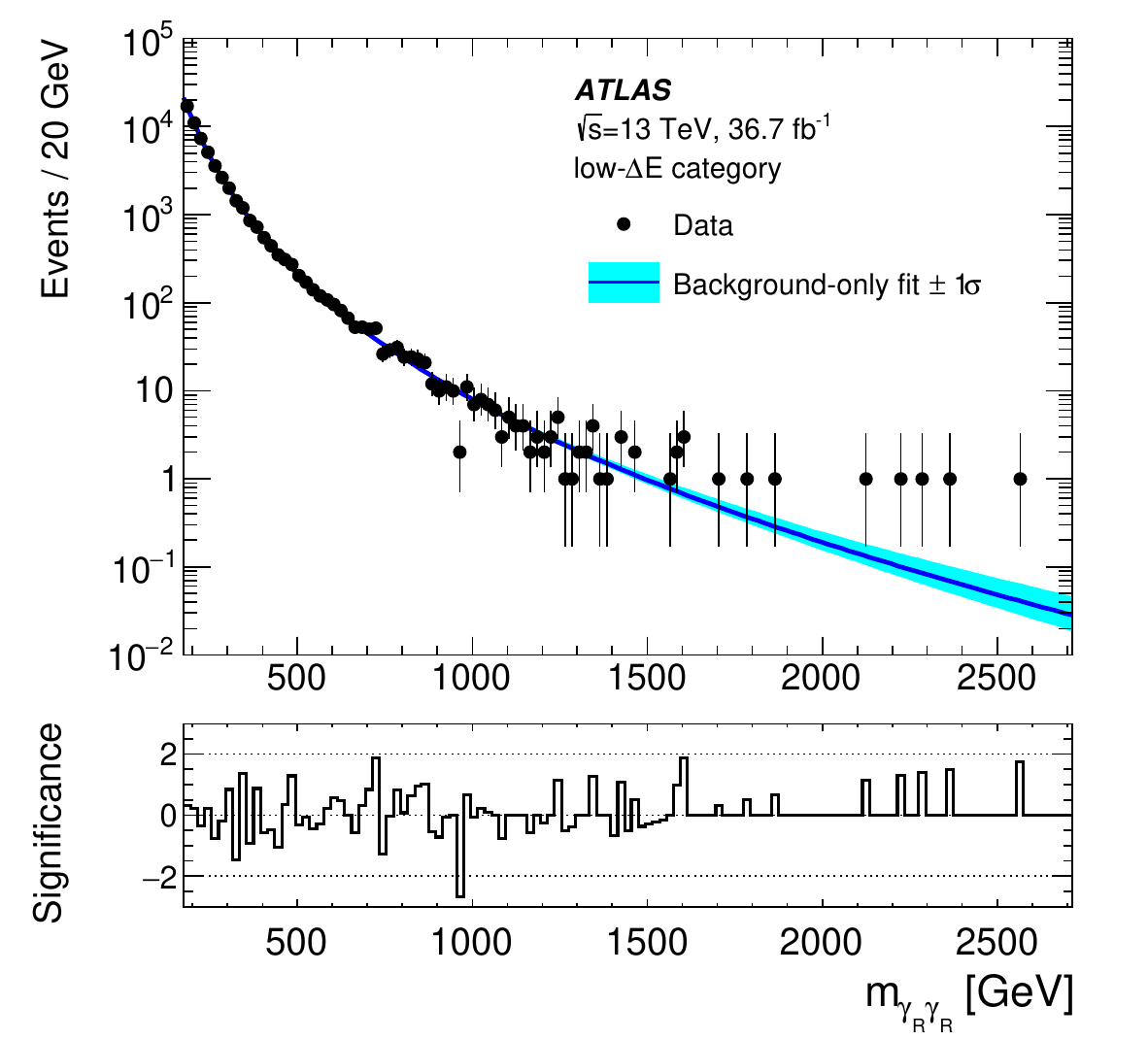}}
	\subfloat{\includegraphics[width=0.5\textwidth]{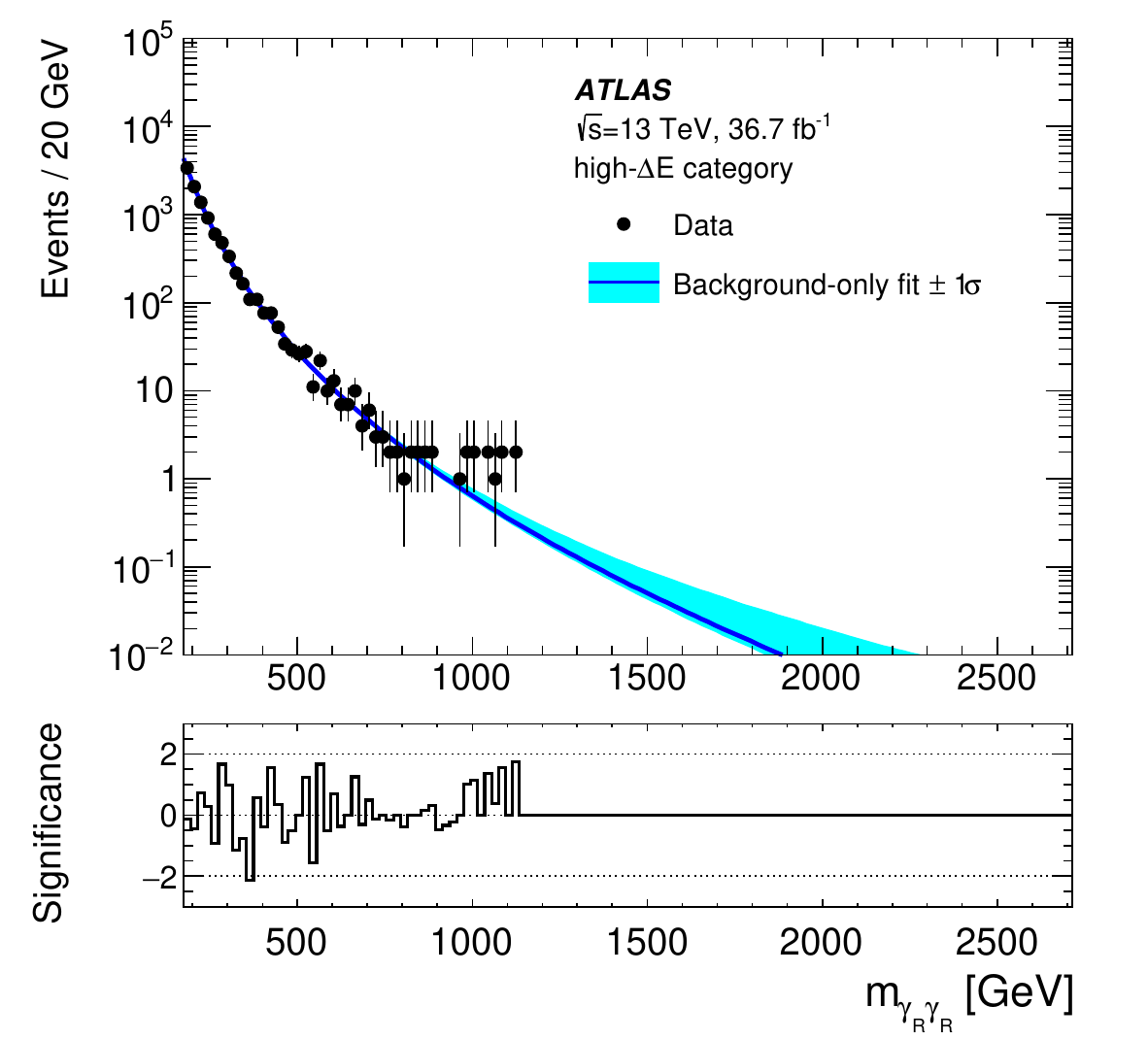}}
	\caption{
	Observed distributions of the mass of two reconstructed photons passing all analysis selections, $\mgg$, for the two signal region categories.
	The background-only fit result is superimposed.
        The $ \pm 1\sigma $ uncertainty originating from the uncertainties in the fit function parameter values is shown as a shaded band around the fit.
		The lower panel of each plot displays the significance associated with the observed event yield in each bin, calculated before considering systematic uncertainties. 
		The calculation assumes that the event yield in each bin is Poisson-distributed with a mean given by the background-only fit. 
		The computation is performed with a one-sided test based on the positive or negative tail of the Poisson distribution, depending on the sign of the difference between the event yield and the fit estimate, with negative significance values quoted for negative differences~\cite{Choudalakis:2011okv}.
    }
	\label{fig:results_fit}
\end{figure}

\begin{figure}[h!]
	\centering
	\subfloat{\includegraphics[width=0.8\textwidth]{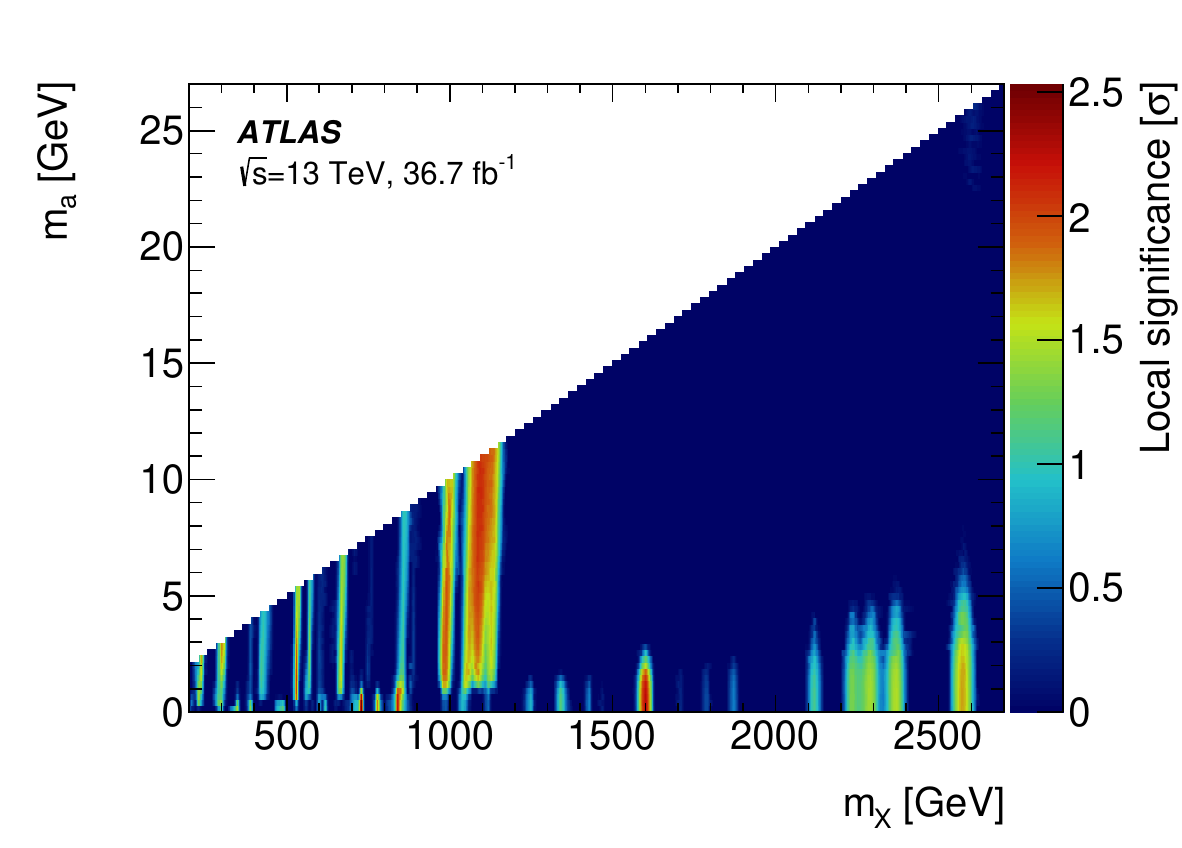}}
	\caption{Results of the search for deviations from the background-only hypothesis in the observed distributions of the $\mgg$, expressed in significance.
	They are presented as a function of $m_{a}$ and $m_{X}$ for the benchmark signal scenario involving a scalar particle $ X $ with narrow width decaying via $X\to aa\to4\gamma$.
	}
	\label{fig:results_significance}
\end{figure}

\begin{figure}[h!]
	\centering
	\subfloat{\includegraphics[width=0.7\textwidth]{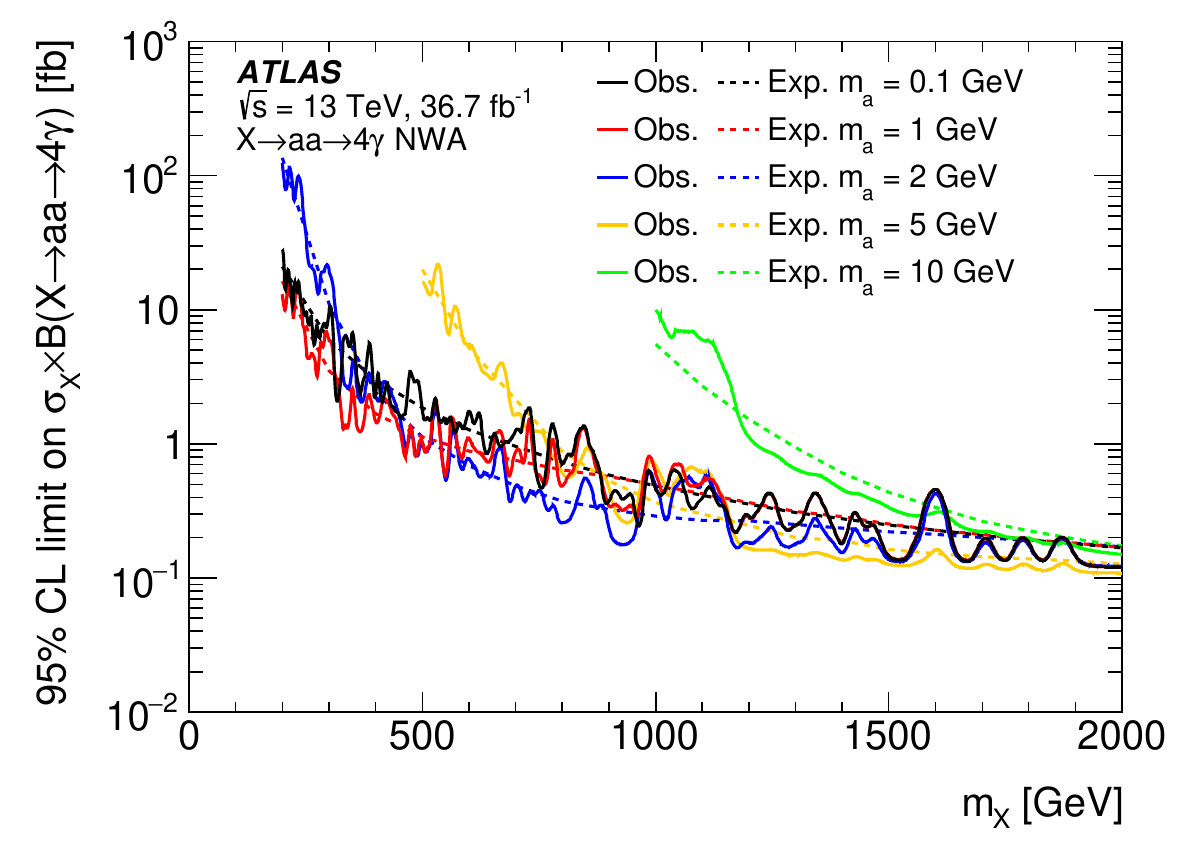}}
	\caption{
	The observed and expected upper limits on the production cross section times the product of branching ratios for the benchmark signal scenario involving a scalar particle $ X $ with narrow width decaying via $X\to aa\to4\gamma$, $\sigma_{X} \times {\mathcal {B}}(X \rightarrow aa) \times {\mathcal {B}}(a \rightarrow \gamma\gamma)^{2}$.
	The limits are calculated using the asymptotic approximation.
	This leads to an underestimate of the limits, especially for $ m_{X}>1~\TeV $ and large $ m_a $.
    The  limits for $m_a = 5$~\GeV\ and 10~\GeV\ do not cover as large a range as the other mass points, since the region of interest is limited to $m_a < 0.01 \times m_X$.
	}
	\label{fig:results_limit_superimposed}
\end{figure}

\begin{figure}[h!]
	\centering
	\subfloat{\includegraphics[width=0.7\textwidth]{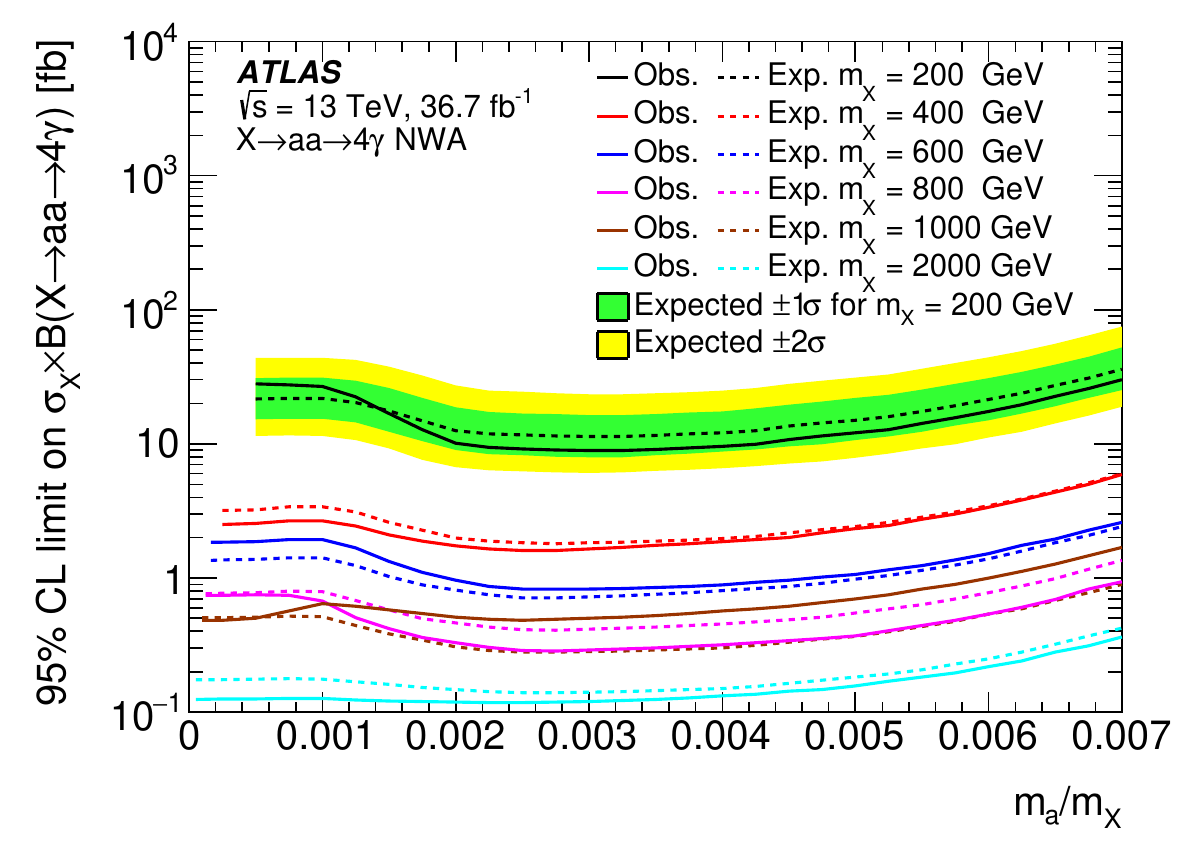}}
	\caption{
	The observed and expected upper limits on the production cross section times the product of branching ratios for the benchmark signal scenario involving a scalar particle $ X $ with narrow width decaying via $X\to aa\to4\gamma$, $\sigma_{X} \times {\mathcal {B}}(X \rightarrow aa) \times {\mathcal {B}}(a \rightarrow \gamma\gamma)^{2}$.
	They are evaluated as a function of $m_a/m_X$ for fixed values of $m_X$.
	The limits are calculated using the asymptotic approximation.
	This leads to an underestimate of the limits, especially for $ m_{X}>1~\TeV $ and large $ m_a $.
	The results for the $ X\to aa\to6\pi^0 $ case are qualitatively similar.
}
	\label{fig:results_limit_superimposed_mamX}
\end{figure}

\begin{figure}[h!]
	\centering
	\subfloat{\includegraphics[width=0.7\textwidth]{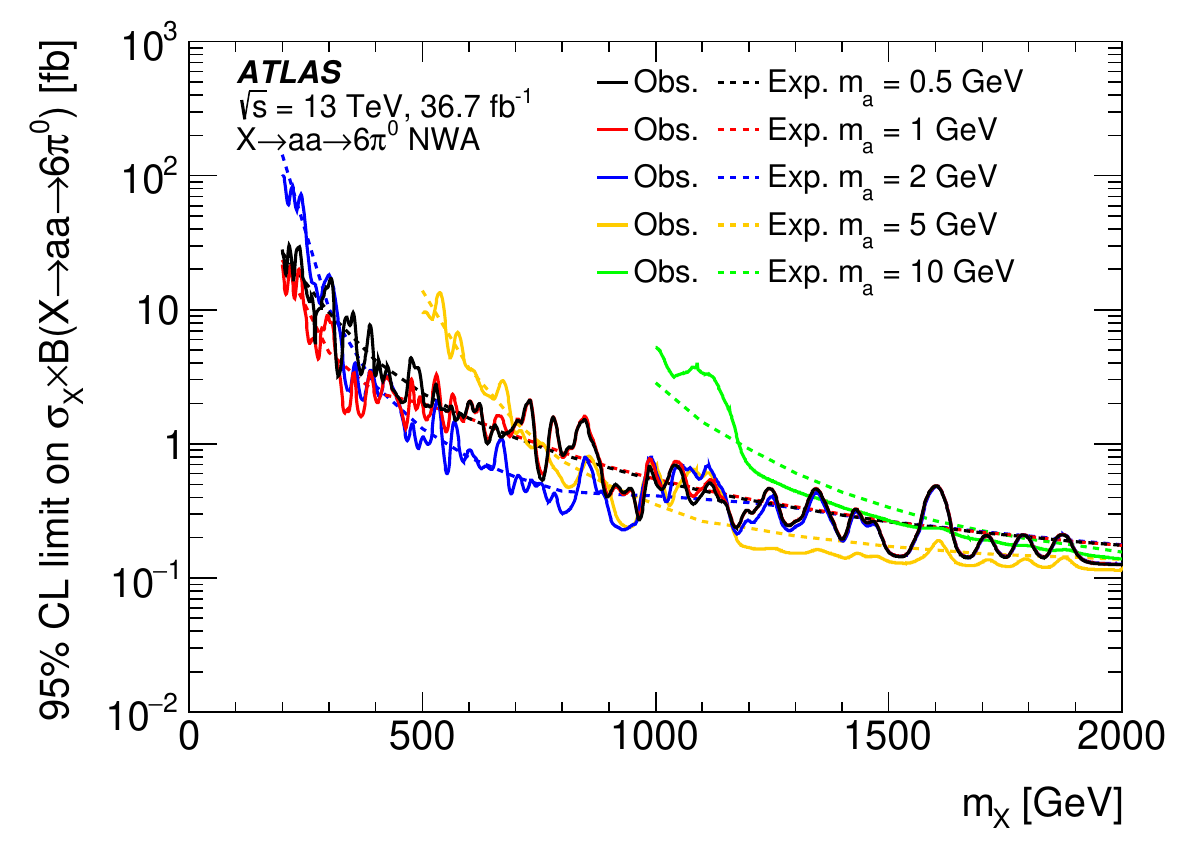}}
	\caption{
	The observed and expected upper limits on the production cross section times the product of branching ratios for the benchmark signal scenario involving a scalar particle $ X $ with narrow width decaying via $X\to aa\to6\pi^0$, $\sigma_{X} \times {\mathcal {B}}(X \rightarrow aa) \times {\mathcal {B}}(a \rightarrow 3\pi^0)^{2}$.
	The limits are calculated using the asymptotic approximation.
	This leads to an underestimate of the limits, especially for $ m_{X}>1~\TeV $ and large $ m_a $.
    The limits for $m_a$ = 5~\GeV\ and 10~\GeV\ do not cover as large a range as the other mass points, since the region of interest is limited to $m_a<0.01\times m_X$.
}
	\label{fig:results_limit_superimposed_sixpion}
\end{figure}

\FloatBarrier
\section{Conclusion} \label{sec:Conclusion}
A search for pairs of highly collimated groupings of photons---photon-jets---that are identified as single, photonlike energy clusters in the EM calorimeter of the ATLAS detector at the LHC is presented.
Data from proton-proton collisions at a center-of-mass energy of 13~\TeV\ collected in 2015 and 2016, corresponding to an integrated luminosity of 36.7~fb$^{-1}$, are used.
Pairs of photon-jets can arise, for example, as the final-state decay products of a new high-mass resonance decaying via new light resonances into highly collimated groupings of photons.
Candidate photon-jet events are initially selected with a loose diphoton trigger and then potential photon-jets are selected using a combination of variables that model EM shower development.
Sensitivity to photon-jets is then increased by categorizing reconstructed photons by one of those shower shape variables and narrow resonances are searched for in the resulting mass distributions of two reconstructed photons.
The observed mass spectra are consistent with the SM background expectation.

The results are interpreted in the context of a BSM scenario containing a high-mass scalar particle with narrow width, $X$, that decays into photon-jets via low-mass intermediate particles with spin 0, $a$.
For the range of $m_X$ investigated, from $ 200~\GeV $ to $ 2~\TeV $, upper limits on $\sigma \times {\mathcal {B}}(X \rightarrow aa) \times {\mathcal {B}}(a \rightarrow \gamma \gamma)^{2}$ are found to range from 0.2~fb to 1~fb over most of the range of $m_X$, for 100~\MeV\ $< m_{a} <$ 2~\GeV, rising to $10$--$100$~fb for values of $m_X$ at the low end of the range, depending upon $m_a$.
Similarly, upper limits on $\sigma \times {\mathcal {B}}(X \rightarrow aa) \times {\mathcal {B}}(a \rightarrow 3\pi^{0})^{2}$ are found to range from 0.2~fb to 1~fb over most of the range of $m_X$, for 500~\MeV\ $< m_{a} <$ 2~\GeV, rising to $10$--$100$~fb for values of $m_X$ at the low end of the range.
These limits are calculated using an asymptotic approximation.
In addition to the calculated upper limits for this benchmark signal scenario, the results, including the evaluation of the observed upper limits, are provided in HepData~\cite{HepData} in a largely model-independent way, to enable reinterpretation in the context of other signal models containing highly collimated photon-jets of low or high photon multiplicity.

\newpage

\section*{Acknowledgments}


We thank CERN for the very successful operation of the LHC, as well as the
support staff from our institutions without whom ATLAS could not be
operated efficiently.

We acknowledge the support of ANPCyT, Argentina; YerPhI, Armenia; ARC, Australia; BMWFW and FWF, Austria; ANAS, Azerbaijan; SSTC, Belarus; CNPq and FAPESP, Brazil; NSERC, NRC and CFI, Canada; CERN; CONICYT, Chile; CAS, MOST and NSFC, China; COLCIENCIAS, Colombia; MSMT CR, MPO CR and VSC CR, Czech Republic; DNRF and DNSRC, Denmark; IN2P3-CNRS, CEA-DRF/IRFU, France; SRNSFG, Georgia; BMBF, HGF, and MPG, Germany; GSRT, Greece; RGC, Hong Kong SAR, China; ISF, I-CORE and Benoziyo Center, Israel; INFN, Italy; MEXT and JSPS, Japan; CNRST, Morocco; NWO, Netherlands; RCN, Norway; MNiSW and NCN, Poland; FCT, Portugal; MNE/IFA, Romania; MES of Russia and NRC KI, Russian Federation; JINR; MESTD, Serbia; MSSR, Slovakia; ARRS and MIZ\v{S}, Slovenia; DST/NRF, South Africa; MINECO, Spain; SRC and Wallenberg Foundation, Sweden; SERI, SNSF and Cantons of Bern and Geneva, Switzerland; MOST, Taiwan; TAEK, Turkey; STFC, United Kingdom; DOE and NSF, United States of America. In addition, individual groups and members have received support from BCKDF, the Canada Council, CANARIE, CRC, Compute Canada, FQRNT, and the Ontario Innovation Trust, Canada; EPLANET, ERC, ERDF, FP7, Horizon 2020 and Marie Sk{\l}odowska-Curie Actions, European Union; Investissements d'Avenir Labex and Idex, ANR, R{\'e}gion Auvergne and Fondation Partager le Savoir, France; DFG and AvH Foundation, Germany; Herakleitos, Thales and Aristeia programmes co-financed by EU-ESF and the Greek NSRF; BSF, GIF and Minerva, Israel; BRF, Norway; CERCA Programme Generalitat de Catalunya, Generalitat Valenciana, Spain; the Royal Society and Leverhulme Trust, United Kingdom.

The crucial computing support from all WLCG partners is acknowledged gratefully, in particular from CERN, the ATLAS Tier-1 facilities at TRIUMF (Canada), NDGF (Denmark, Norway, Sweden), CC-IN2P3 (France), KIT/GridKA (Germany), INFN-CNAF (Italy), NL-T1 (Netherlands), PIC (Spain), ASGC (Taiwan), RAL (UK) and BNL (USA), the Tier-2 facilities worldwide and large non-WLCG resource providers. Major contributors of computing resources are listed in Ref.~\cite{ATL-GEN-PUB-2016-002}.

\printbibliography

\clearpage 
 
\begin{flushleft}
{\Large The ATLAS Collaboration}

\bigskip

M.~Aaboud$^\textrm{\scriptsize 34d}$,
G.~Aad$^\textrm{\scriptsize 99}$,
B.~Abbott$^\textrm{\scriptsize 124}$,
O.~Abdinov$^\textrm{\scriptsize 13,*}$,
B.~Abeloos$^\textrm{\scriptsize 128}$,
D.K.~Abhayasinghe$^\textrm{\scriptsize 91}$,
S.H.~Abidi$^\textrm{\scriptsize 164}$,
O.S.~AbouZeid$^\textrm{\scriptsize 143}$,
N.L.~Abraham$^\textrm{\scriptsize 153}$,
H.~Abramowicz$^\textrm{\scriptsize 158}$
% ...
\bigskip
\\

% ...
$^{13}$Institute of Physics, Azerbaijan Academy of Sciences, Baku; Azerbaijan.\\
% ...
$^{34}$$^{(a)}$Facult\'e des Sciences Ain Chock, R\'eseau Universitaire de Physique des Hautes Energies - Universit\'e Hassan II, Casablanca;$^{(b)}$Centre National de l'Energie des Sciences Techniques Nucleaires (CNESTEN), Rabat;$^{(c)}$Facult\'e des Sciences Semlalia, Universit\'e Cadi Ayyad, LPHEA-Marrakech;$^{(d)}$Facult\'e des Sciences, Universit\'e Mohamed Premier and LPTPM, Oujda;$^{(e)}$Facult\'e des sciences, Universit\'e Mohammed V, Rabat; Morocco.\\
% ...
$^{91}$Department of Physics, Royal Holloway University of London, Egham; United Kingdom.\\
% ...
$^{99}$CPPM, Aix-Marseille Universit\'e, CNRS/IN2P3, Marseille; France.\\
% ...
$^{124}$Homer L. Dodge Department of Physics and Astronomy, University of Oklahoma, Norman OK; United States of America.\\
% ...
$^{128}$LAL, Universit\'e Paris-Sud, CNRS/IN2P3, Universit\'e Paris-Saclay, Orsay; France.\\
% ...
$^{143}$Santa Cruz Institute for Particle Physics, University of California Santa Cruz, Santa Cruz CA; United States of America.\\
% ...
$^{153}$Department of Physics and Astronomy, University of Sussex, Brighton; United Kingdom.\\
% ...
$^{158}$Raymond and Beverly Sackler School of Physics and Astronomy, Tel Aviv University, Tel Aviv; Israel.\\
% ...
$^{164}$Department of Physics, University of Toronto, Toronto ON; Canada.\\
% ...
$^{a}$ Also at Borough of Manhattan Community College, City University of New York, New York City; United States of America.\\
$^{b}$ Also at Centre for High Performance Computing, CSIR Campus, Rosebank, Cape Town; South Africa.\\
$^{c}$ Also at CERN, Geneva; Switzerland.\\
$^{d}$ Also at CPPM, Aix-Marseille Universit\'e, CNRS/IN2P3, Marseille; France.\\
$^{e}$ Also at Departament de Fisica de la Universitat Autonoma de Barcelona, Barcelona; Spain.\\
% ...
$^{*}$ Deceased

\end{flushleft}


\end{document}